\documentclass[12pt]{article}

\RequirePackage[sorting=none,
                style=nature,
                backend=biber]{biblatex}
\RequirePackage{times}
\RequirePackage{fullpage}
\RequirePackage{ifthen}

\newcommand{\spacing}[1]{\renewcommand{\baselinestretch}{#1}\large\normalsize}
\spacing{2}

\makeatletter
\def\@maketitle{%
  \newpage\spacing{1}\setlength{\parskip}{12pt}%
    {\Large\bfseries\noindent\sloppy \textsf{\@title} \par}%
    {\noindent\sloppy \@author}%
}

\newenvironment{affiliations}{%
    \setcounter{enumi}{1}%
    \setlength{\parindent}{0in}%
    \slshape\sloppy%
    \begin{list}{\upshape$^{\arabic{enumi}}$}{%
        \usecounter{enumi}%
        \setlength{\leftmargin}{0in}%
        \setlength{\topsep}{0in}%
        \setlength{\labelsep}{0in}%
        \setlength{\labelwidth}{0in}%
        \setlength{\listparindent}{0in}%
        \setlength{\itemsep}{0ex}%
        \setlength{\parsep}{0in}%
        }
    }{\end{list}\par\vspace{12pt}}

\renewenvironment{abstract}{%
    \setlength{\parindent}{0in}%
    \setlength{\parskip}{0in}%
    \bfseries%
    }{\par\vspace{-6pt}}

\renewcommand{\section}{\@startsection {section}{1}{0pt}%
    {-6pt}{1pt}%
    {\bfseries}%
    }
    
\renewcommand{\subsection}{\@startsection {subsection}{2}{0pt}%
    {-0pt}{-0.5em}%
    {\bfseries}*%
    }

\newenvironment{methods}{%
    \section*{Methods}%
    \setlength{\parskip}{12pt}%
    }{}

\newenvironment{addendum}{%
    \setlength{\parindent}{0in}%
    \small%
    \begin{list}{Acknowledgements}{%
        \setlength{\leftmargin}{0in}%
        \setlength{\listparindent}{0in}%
        \setlength{\labelsep}{0em}%
        \setlength{\labelwidth}{0in}%
        \setlength{\itemsep}{12pt}%
        }
    }
    {\end{list}\normalsize}

\makeatother

\usepackage{xcolor}
\usepackage{graphicx}
\usepackage{float}
\usepackage{newfloat}
\DeclareFloatingEnvironment[name={Extended Data Figure}]{extdatafig}
\DeclareFloatingEnvironment[name={Supplementary Figure}]{suppfig}
\usepackage{caption}
\DeclareCaptionLabelSeparator{bar}{ $|$ }
\usepackage[export]{adjustbox}
\usepackage{siunitx}
\usepackage{sidecap}
\sidecaptionvpos{figure}{c}
\usepackage{tablefootnote}
\usepackage{threeparttable}
\usepackage{threeparttablex}
\usepackage{longtable}
\usepackage{amssymb}
\usepackage{prettyref}
\usepackage{hyperref}
\usepackage{filecontents}

\makeatletter
\AtEveryCitekey{%
  \ifcsundef{blx@entry@refsegment@\the\c@refsection @\thefield{entrykey}}
    {\csnumgdef{blx@entry@refsegment@\the\c@refsection @\thefield{entrykey}}{\the\c@refsegment}}
    {}}
\defbibcheck{onlynew}{%
  \ifnumless{0\csuse{blx@entry@refsegment@\the\c@refsection @\thefield{entrykey}}}{\the\c@refsegment}
    {\skipentry}
    {}}
\makeatother

\newrefformat{sitab}{Supplementary Table \ref{#1}}
\newrefformat{fig}{Figure \ref{#1}}
\newrefformat{sifig}{Supplementary Figure \ref{#1}}
\newrefformat{edfig}{Extended Data Figure \ref{#1}}

\AtEveryBibitem{\clearfield{month}}
\AtEveryCitekey{\clearfield{month}}

\title{An extremely powerful long lived superluminal ejection from the black hole MAXI J1820$+$070}

\author{J. S. Bright$^1$, R. P. Fender$^{1,2}$, S. E. Motta$^1$, D. R. A. Williams$^1$, J. Moldon$^{3,4}$, R. M. Plotkin$^{5,6}$, J. C. A. Miller-Jones$^6$, I. Heywood$^{1,7,8}$, E. Tremou$^9$, R. Beswick$^4$, G. R. Sivakoff$^{10}$, S. Corbel$^{9,11}$, D. A. H. Buckley$^{12}$, J. Homan$^{13,14,15}$, E. Gallo$^{16}$, A. J. Tetarenko$^{17}$, T. D. Russell$^{18}$, D. A. Green$^{19}$, D. Titterington$^{19}$, P. A. Woudt$^{2,20}$, R. P. Armstrong$^{2,1,8}$, P. J. Groot$^{21,2,12}$, A. Horesh$^{22}$, A. J. van der Horst$^{23,24}$, E. G. K{\"o}rding$^{21}$, V. A. McBride$^{2,12,25}$, A. Rowlinson$^{18,26}$ \& R. A. M. J. Wijers$^{18}$}

\begin{document}

\maketitle

\begin{affiliations}
 \item Astrophysics, Department of Physics, University of Oxford, Keble Road, Oxford, OX1 3RH, UK
  \item Department of Astronomy, University of Cape Town, Private Bag X3, Rondenbosch, 7701, South Africa
 \item Instituto de Astrof\'isica de Andaluc\'ia (IAA, CSIC), Glorieta de las Astronom\'ia, s/n, E-18008 Granada, Spain
 \item Jodrell Bank Centre for Astrophysics, The University of Manchester, M13 9PL, UK
 \item Department of Physics, University of Nevada, Reno, Nevada, 89557, USA
 \item International Centre for Radio Astronomy Research, Curtin University, GPO Box, U1987, Perth, WA 6845, Australia
 \item Department of Physics and Electronics, Rhodes University, PO Box 94, Grahamstown, 6140, South Africa
 \item South African Radio Astronomy Observatory (SARAO), 2 Fir Street, Observatory, Cape Town, 7925, South Africa
 \item AIM/CEA Paris-Saclay, Universit{\`e} Paris Diderot, CNRS, F-91191 Gif-sur-Yvette, France
 \item Department of Physics, CCIS 4-183, University of Alberta, Edmonton, AB, T6G 2E1, Canada
 \item Station de Radioastronomie de Nan\c{c}ay, Observatoire de Paris, PSL Research University, CNRS, Univ. Orl{\'e}ans, 18330 Nan\c{c}ay, France
 \item South African Astronomical Observatory, P.O. Box 9, Observatory 7935, Cape Town, South Africa
 \item Eurika Scientific, Inc., 2452 Delmer Street, Oakland, CA 94602, USA
 \item SRON, Netherlands Institute for Space Research, Sorbonnelaan 2, 3584 CA Utrecht, The Netherlands
 \item MIT Kavli Institute for Astrophysics and Space Research, 70 Vassar Street, Cambridge, MA 02139, USA
 \item Department of Astronomy, University of Michigan, 1085 S University, Ann Arbor, MI 48109, USA
 \item East Asian Observatory, 660 N. A'oh\={o}k\={u} Place, University Park, Hilo, Hawaii, 96720, USA
 \item Anton Pannekoek Institute, University of Amsterdam, Postbus 94294, 1090 GE, Amsterdam, The Netherlands
 \item Astrophysics Group, Cavendish Laboratory, 19 J. J. Thomson Avenue, Cambridge, CB3 OHE, UK
 \item Inter-University Institute of Data Intensive Astronomy, Department of Astronomy, University of Cape Town, Private Bag X3, Rondebosch 7701, South Africa
 \item Department of Astrophysics/IMAPP, Radboud University Nijmegen, P.O. Box 9010, 6500 GL, Nijmegen, The Netherlands
 \item Racah Institute of Physics, The Hebrew University of Jerusalem, Jerusalem 91904, Israel
 \item Department of Physics, the George Washington University, 725 21st Street NW, Washington, DC 20052, USA
 \item Astronomy, Physics and Statistics Institute of Sciences (APSIS), 725 21st Street NW, Washington, DC 20052, USA
 \item IAU Office of Astronomy for Development, Cape Town, South Africa
 \item Netherlands Institute for Radio Astronomy (ASTRON), Oude Hoogeveensedijk 4, 7991 PD, Dwingeloo, The Netherlands
 
\end{affiliations}

\clearpage

\newrefsegment

\begin{abstract}
Black holes in binary systems execute patterns of outburst activity where two characteristic X-ray states are associated with different behaviours observed at radio wavelengths. The hard state is associated with radio emission indicative of a continuously replenished, collimated, relativistic jet, whereas the soft state is rarely associated with radio emission, and never continuously, implying the absence of a quasi-steady jet. Here we report radio observations of the black hole transient MAXI J1820$+$070 during its 2018 outburst. As the black hole transitioned from the hard to soft state we observed an isolated radio flare, which, using high angular resolution radio observations, we connect with the launch of bi-polar relativistic ejecta. This flare occurs as the radio emission of the core jet is suppressed by a factor of over 800. We monitor the evolution of the ejecta over 200 days and to a maximum separation of 10$''$, during which period it remains detectable due to in-situ particle acceleration. Using simultaneous radio observations sensitive to different angular scales we calculate an accurate estimate of energy content of the approaching ejection. This energy estimate is far larger than that derived from state transition radio flare, suggesting a systematic underestimate of jet energetics.
\end{abstract}

Black hole X-ray binary (BHXRB) systems consist of a stellar-mass black hole accreting material via Roche lobe overflow from a main sequence companion star. X-ray observations of such systems, which probe their accretion flow, have revealed the existence of two primary accretion states, termed hard and soft\supercite{fender2004, corbel2004}. In the hard state the X-ray spectrum is non-thermal, and thought to be dominated by emission from an inner accretion disk corona. In the soft state coronal emission is suppressed, and the X-ray spectrum is well described by thermal emission from the accretion disk itself. Contemporaneous radio observations, which probe the jets, show that the accretion state of a BHXRB system determines the form of the outflows it produces\supercite{blandford1997,blandford1982,rees1982,fender2004,corbel2004}. During the hard state radio emission is from a flat spectrum, collimated, compact (solar-system scale) jet\supercite{stirling2001,dhawan2000} which is quenched in the soft state\supercite{russell2011,coriat2011,rushton2016,russell2019}. The most dramatic outburst behaviour occurs as sources transition from the hard to the soft accretion state. During the transition, as the core jet quenches, systems exhibit short timescale (of the order hours) radio flaring superposed on the decaying core jet flux \supercite{fender2004}. These flares have been associated with the ejection of discrete (apparently no longer connected spatially to the black hole) knots of material, which can be observed to move (sometimes apparently superluminally) away from the black hole, reaching separations tens of thousands times farther than that of the core jet \supercite{mirabel1994}. The mechanism(s) causing the launch of these ejections, as well as the radio flaring, are not well understood. Jets and ejections represent two of the primary channels through which galactic black holes return matter and energy into their surroundings and studying them is key to understanding feedback processes and their effects on the environment from black holes over a range of mass scales.

MAXI J1820+070/ASASSN-18ey\supercite{tucker2018,kawamuro2018,uttley2018,torres2019} was discovered at optical wavelengths by the All-Sky Automated Survey for SuperNovae (ASAS-SN\supercite{shappee2014}) project on 7\textsuperscript{th} March 2018 (MJD 58184), and around six days later in X-rays by the Monitor of All-sky X-ray Image (MAXI\supercite{kawamuro2018}). Soon after, it was classified as a candidate black hole X-ray binary (BHXRB) by the Neutron Star Interior Composition Explorer (NICER) based on its timing properties\supercite{uttley2018} (and later dynamically confirmed from its mass function\supercite{torres2019}). The Neil Gehrels Swift Observatory (\textit{Swift}) Burst Alert Telescope (BAT) triggered on J1820 on MJD 58189\supercite{kennea2018}, prompting rapid robotic follow-up\supercite{staley2013} with the Arcminute Microkelvin Imager Large Array (AMI-LA) only 90 minutes later, the earliest radio detection of the outburst of a new black hole X-ray binary ever reported. The relatively close proximity\supercite{gandhi2019} ($3.8^{+2.9}_{-1.2}$ kpc) and brightness of J1820 allowed for extremely good coverage of the outburst across the electromagnetic spectrum\supercite{tucker2018,shidatsu2018,shidatsu2019,munozdarias2019,kajava2019,kara2019}.

\section*{Results}

Throughout its outburst we monitored J1820 intensively with a range of radio telescopes: the AMI-LA, Multi-Element Radio Linked Interferometer Network (eMERLIN), Meer Karoo Array Telescope (MeerKAT), the Karl G. Jansky Very Large Array (VLA) and the Very Long Baseline Array (VLBA; ordered in terms of time spent on source; see the Methods section for the details of our observations) as well as at X-ray wavelengths with \textit{Swift}. In \prettyref{fig:lxlr} we present a subset of our over $200\,\textrm{d}$ of $15.5\,\textrm{GHz}$ radio monitoring as a function of X-ray luminosity, selecting only (quasi-)simultaneous observations. This radio--X-ray plane is typically used to study the non-linear correlation between radio and X-ray fluxes from black holes in the hard state, from which radio emission is always detected, revealing the connection between accretion rate and jet power. Sources in the soft state, which corresponds to dramatically different accretion properties with respect to the hard state (see Supplementary Information), are rarely detected in the radio band and consequently are not usually shown on the diagram. Our (quasi-)simultaneous radio and X-ray coverage includes the hard states at the start and end of the outburst (before and after MJD 58303.5 and 58393, respectively), the entirety of the soft state (between MJD 58310.7 and 58380) and the intermediate states, where the relative X-ray emission contribution of the disk and corona are evolving quickly and the core radio jet is quenching or restarting (MJD 58303.5 to 58310.7, and MJD 58380 to 58393, respectively)\supercite{shidatsu2019}. During the decay phase (where the radio flux from the quenching core jet is dropping) after the initial hard state (i.e. during the first intermediate state) we observe a `flare' event, characterised by a  short timescale rise and decay ($\sim\textrm{12}\,\textrm{h}$ in total) of the radio emission from J1820 (see \prettyref{edfig:amiflare}). These events are thought to be caused not by the compact core jet, but instead by discrete relativistic ejections\supercite{fender2009}. As expected, J1820 is detected throughout the hard state, as well as during the decay and rise (where the compact jet is now switching back on) phases before and after the soft state, respectively. The radio--X-ray correlation indicates that the source is `radio loud' in the sense that it lies on the higher track in the plane, similarly to the archetypal source GX 339$-$4\supercite{corbel2013}.\\

\begin{SCfigure}
\includegraphics[width=0.545\textwidth]{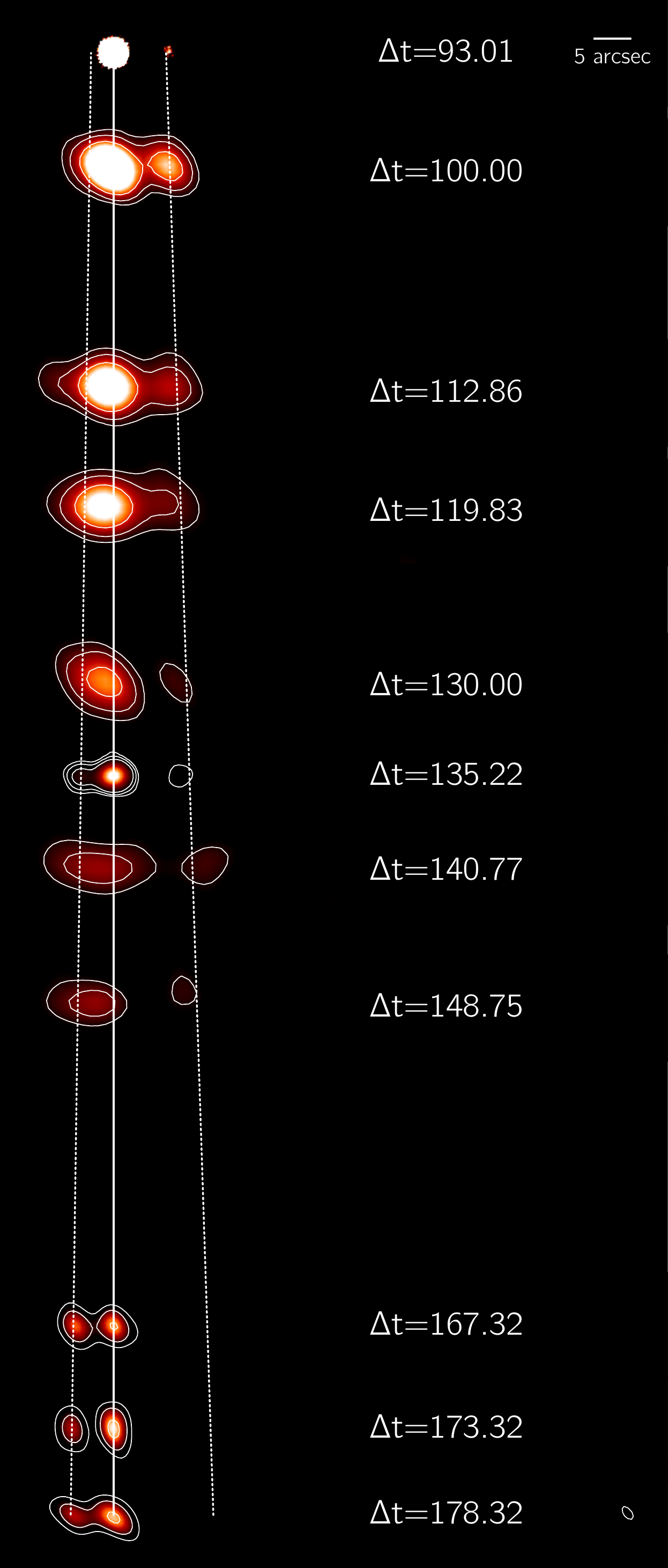}
\caption{\textbf{A subset of our resolved images of the core and ejections from J1820.} A subset of images of J1820 from eMERLIN ($\Delta\textrm{t}=\textrm{93.01}$) MeerKAT ($\Delta\textrm{t}=\textrm{100.00}$, $\textrm{112.86}$, $\textrm{119.83}$, $\textrm{130.00}$ and $\textrm{140.77}$, $\textrm{148.75}$) and the VLA ($\Delta\textrm{t}=\textrm{135.22}$, $\textrm{167.32}$, $\textrm{173.32}$ and $\textrm{178.32}$) where we resolve at least one ejecta from the core. All images have been rotated $\sim\textrm{65}^{\circ}$ anticlockwise. $\Delta\textrm{t}$ is the time, in days, since the start of the radio flare that occurred during the hard to soft state transition ($\Delta\textrm{t}=0$ at MJD 58305.68), and is shown to the right of each observation. All observations are shown with the same angular scale, and a scale bar is shown in the top right of the figure. For the MeerKAT observations, contours show $\textrm{40}\,\mu\textrm{Jy}\,\textrm{beam}^{-\mathrm{1}}\times(\sqrt{\textrm{2}})^{n}$ for $n=4,6,8$ and the colour-scale is linear between $\textrm{0.1}\,\textrm{mJy}\,\textrm{beam}^{-\mathrm{1}}$ and $\textrm{1}\,\textrm{mJy}\,\textrm{beam}^{-\mathrm{1}}$. For the VLA observations, contours show $\textrm{8}\,\mu\textrm{Jy}\,\textrm{beam}^{-\mathrm{1}}\times(\sqrt{\textrm{2}})^{n}$ for $n=4,6,8$, and the colour-scale is linear between $\textrm{0.05}\,\textrm{mJy}\,\textrm{beam}^{-\mathrm{1}}$ and $\textrm{0.15}\,\textrm{mJy}\,\textrm{beam}^{-\mathrm{1}}$ for all but the first observation, which shares the same scale as the MeerKAT data. The colour-scale for the eMERLIN observation is linear between $\textrm{0.2}\,\textrm{mJy}\,\textrm{beam}^{-\mathrm{1}}$ and $\textrm{0.3}\,\textrm{mJy}\,\textrm{beam}^{-\mathrm{1}}$. The white vertical solid line marks the position of the core, determined from hard state observations made with eMERLIN. The right and left dashed lines show the best fit ballistic trajectory of approaching and receding ejection components, respectively. These fits are constrained by the observations presented in \prettyref{sitab:allpos}. Details on the data reduction procedure are presented in the Methods section.\label{fig:other}}
\end{SCfigure}

Remarkably, however, J1820 is also detected continuously, at a lower level, in 56 observations throughout its $\textrm{80}\,\textrm{d}$ soft state (which are also demonstrated as a function of X-ray luminosity in \prettyref{fig:lxlr}). In the most comprehensive previous ensemble study of radio observations of the BHXRB soft state\supercite{fender2009} it was demonstrated that for all but one source (XTE J1748$-$288\supercite{brocksopp2007}) there is either no radio emission, or only transient emission, during the soft state, and for that one exception the nature of the emission is very poorly determined. Since then only one source, MAXI J1535$-$571, has shown such long lived soft state emission\supercite{russell2019}. We show in \prettyref{fig:lxlr} the previous deepest limits on radio emission in the soft state from other BHXRBs. Based on the AMI-LA radio flux density monitoring alone, the nature of the soft state radio emission from J1820 is unclear. Without high-resolution radio images the continued soft state emission could, in principle, be interpreted as evidence for a causal connection between the accretion flow and radio emission, i.e. ongoing core jet production in the soft state. A radio image with the VLBA (\prettyref{edfig:vlbi}), at $\textrm{15}\,\textrm{GHz}$, $\sim3\,\textrm{d}$ into the intermediate state as J1820 transitioned from the hard to soft state reveals, however, that the core is not detected to a $\textrm{3}\sigma$ flux density limit of $\sim\textrm{420}\,\mu\textrm{Jy}\,\textrm{beam}^{-\mathrm{1}}$, and that there are components that can be associated with relativistic ejections both approaching and receding from the Earth. Images with eMERLIN (\prettyref{fig:em}) over the following $\sim3\,\textrm{w}$ show the approaching component moving away from the black hole. Further radio images with the eMERLIN, MeerKAT and VLA radio telescopes (\prettyref{fig:other}), when J1820 had returned to the hard X-ray state, reveal that the approaching jet is still detected over $\textrm{140}\,\textrm{d}$ after J1820 transitioned to the soft state. At later times a receding ejection is resolved with MeerKAT and the VLA, which we observe up to $\textrm{175}\,\textrm{d}$ after the start of the soft state. A comparison between the time evolution of the (frequency-scaled) flux density from the resolved approaching ejection and the (unresolved) flux density measured by the AMI-LA during the soft state reveals that all of the radio emission can be attributed to this ejection (\prettyref{fig:lc}). The flux density from the approaching ejection shows multiple decay rates. After a short rise, the flux density decays with a timescale (e-folding time) of $\sim\textrm{6}\,\textrm{d}$. However, after $\sim\textrm{10}\,\textrm{d}$ it re-brightens and undergoes a much slower decay with a timescale of $\sim\textrm{50}\,\textrm{d}$, transitioning $\sim\textrm{60}\,\textrm{d}$ later to a faster decay rate with a $\sim\textrm{20}\,\textrm{d}$ timescale (see Supplementary Information and \prettyref{edfig:decayfits}).

When fitting for the proper motion (angular velocity across the sky) of the receding and approaching ejection components we consider both a ballistic and constant deceleration model, finding linear proper motions of $\mu_{\textrm{app}}=\textrm{77}\pm\textrm{1}\,\textrm{mas}\,\textrm{d}^{\mathrm{-1}}$ and $\mu_{\textrm{rec}}=\textrm{33}\pm\textrm{1}\,\textrm{mas}\,\textrm{d}^{\mathrm{-1}}$, and initial velocities of $\mu_{\textrm{app}}=\textrm{101}\pm\textrm{3}\,\textrm{mas}\,\textrm{d}^{\mathrm{-1}}$ and $\mu_{\textrm{rec}}=\textrm{58}\pm\textrm{6}\,\textrm{mas}\,\textrm{d}^{\mathrm{-1}}$ for a constant deceleration (see \prettyref{edfig:propermotion}). These are among the highest proper motions ever measured from an astronomical object outside of the Solar System. Our linear proper motion model corresponds to apparent transverse velocities of at least $\textrm{1.7}\,c$ and $\textrm{0.7}\,c$ (for the approaching and receding components, respectively) at a distance of $\textrm{3.8}\,\textrm{kpc}$. J1820, therefore, becomes one of a small number of black hole binaries to have produced jets with apparent superluminal motion\supercite{mirabel1994,hjellming1995,hannikainen2009,russell2019}. The launch date for both models is consistent with MJD 58306, coinciding with the radio flare observed with the AMI-LA during the intermediate state, as the source moved from the hard to the soft state. The proper motion of each ejection can be independently related to its velocity, ejection angle to the observer's line of sight, and the distance to the source from the observer. From a combination of the approaching and receding proper motions (from the linear fit) we can calculate the product $\beta\textrm{cos}\theta=\textrm{0.40}\pm\textrm{0.02}$ (where $\beta$ is the ejection's velocity in units of $c$, and $\theta$ is the jet inclination angle), a quantity that is independent of distance and assumes symmetric ejections with the same speeds\supercite{mirabel1994}. From this we constrain a maximum angle to the line of sight of $\textrm{66}^{\circ}\pm\textrm{1}^{\circ}$ (for $\beta=1$). We can also calculate a maximum distance to the source of $\textrm{3.5}\pm\textrm{0.2}\,\textrm{kpc}$. This corresponds to the distance beyond which a more extreme angle to the line of sight than our calculated maximum angle would be required to explain the observed proper motions. For the constant deceleration model we find a maximum angle to the line of sight of $\textrm{74}^{\circ}\pm\textrm{2}^{\circ}$ and a maximum distance of $\textrm{2.3}\pm\textrm{0.6}\,\textrm{kpc}$, respectively. However, a measured radio parallax distance to J1820 of $\textrm{3.1}\pm\textrm{0.3}\,\textrm{kpc}$ rules out the deceleration model\supercite{atri2019}. The uncertainty in distance, combined with a significantly relativistic jet, means that we can only place a lower limit on its bulk Lorentz factor of $\Gamma>\textrm{1.7}$ (the apparent velocity, corresponding to  $v>0.8\,c$)\supercite{fender2003}.

On MJD $\textrm{58396}$ ($\Delta\textrm{t}=\textrm{91.02}$) and $\textrm{58398}$ ($\Delta\textrm{t}=\textrm{93.01}$) we observed J1820 with MeerKAT and eMERLIN, respectively, at very similar frequencies ($1.28\,\textrm{GHz}$ and $1.51\,\textrm{GHz}$). These telescopes probe very different angular scales, with synthesised beams of $\textrm{7.9}''\times\textrm{5.4}''$ and $\textrm{0.31}''\times\textrm{0.2}''$, respectively. In both observations the approaching jet component is detected. The flux density measured by MeerKAT is around $\textrm{2}\,\textrm{mJy}$, approximately 85\% of which is resolved out at the angular scales probed by eMERLIN (which, due to its longer baselines, is not sensitive to structure on the angular scales probed by MeerKAT and thus recovers only $\textrm{0.3}\,\textrm{mJy}$). Although these observations were not taken strictly simultaneously, the time difference between the two observations is likely not enough to account for this large discrepancy given the observed decay rate (see Supplementary Information). Taking the minimum angular size probed by each observation and the radio parallax distance allows us to set a range of sizes that the $\sim\textrm{85}$\% resolved out flux density ($\sim\textrm{1.7}\,\textrm{mJy}$) is emitted from: between $\sim\textrm{6.2}\times\textrm{10}^{\mathrm{2}}\,\textrm{AU}$ and $\sim\textrm{1.7}\times\textrm{10}^{\mathrm{4}}\,\textrm{AU}$. The emitting region size is the most important measurement for estimating the internal energy of a synchrotron-emitting plasma (which would be significantly underestimated by the integrated radiative power output over our observing campaign). Using our physical size constraints we calculate\supercite{longairhea} lower limits to the internal energy in the range $\textrm{2.1}\times\textrm{10}^{\mathrm{41}}\,\textrm{erg}<\textrm{E}_{\mathrm{f}}<\textrm{1.5}\times\textrm{10}^{\mathrm{43}}\,\textrm{erg}$ at the time of these near simultaneous observations. This also allows us to constrain the equipartition magnetic field corresponding to this range of energies to be between $\textrm{4.9}\times\textrm{10}^{\mathrm{-5}}\,\textrm{G}$ and $\textrm{8.3}\times\textrm{10}^{\mathrm{-4}}\,\textrm{G}$. Our derived lower limit to the minimum energy is orders of magnitude larger than the internal energy associated with the radio flare ($\textrm{E}_{\mathrm{i}}$; thought to be a signature of the launch of transient ejections) observed during the hard to soft state transition\supercite{bright2018}. This flare had an associated minimum internal energy of $\textrm{E}_{\mathrm{i}}\sim\textrm{2}\times\textrm{10}^{\mathrm{37}}\,\textrm{erg}$. This estimate assumes that the flare is the result of the launch of an expanding plasmoid (synchrotron emitting plasma), with the peak caused by an optical depth transition from thick to thin\supercite{vanderlaan1966,tetarenko2017,fender2019}. A larger internal energy estimate can be derived from the peak flux density of the flare ($\sim\textrm{46}\,\textrm{mJy}$) and its rise time ($\sim\textrm{6.7}\,\textrm{hours}$), giving $\textrm{E}_{\mathrm{i}}=\textrm{2}\times\textrm{10}^{\mathrm{39}}\,\textrm{erg}$, where we assume an expansion speed of $\textrm{0.05}\,c$ (There is strong evidence\supercite{tetarenko2017,fender2019,russell2019} to suggest that these ejections expand significantly slower than $c$). However, this larger estimate for the flare's energetics is still two orders of magnitude below the estimate of the ejecta energetics.

\section*{Discussion}

Persistent, slowly evolving radio emission from moving relativistic ejections has been observed in three XRB systems (XTE J1550$-$564, H1743$-$322 and MAXI J1535$-$571) previously. In XTE J1550$-$564, dynamic ejections were observed on small ($<300\,\textrm{mas}$) angular scales following a radio flare\supercite{hannikainen2001}. These ejections then went `dark', and were detected again over two years later due to a re-brightening episode thought to be the result of an interaction with the wall of an ISM density cavity\supercite{corbel2002,tomsick2003,migliori2017}. A similar explanation has been invoked to explain the large scale jets in H1743$-$322\supercite{corbel2005}. In MAXI J1535$-$571 the approaching ejection was tracked for $\sim\textrm{300}\,\mathrm{d}$, after being detected for the first time $\sim\textrm{90}\,\mathrm{d}$ after its inferred launch date\supercite{russell2019}. The ejection was not resolved at an angular separation from the core of less than $4''$, but was tracked out to over $15''$. This allowed the launch time to be constrained to a $\sim\mathrm{5}\,\mathrm{d}$ window, consistent with occurring just before a radio flaring event (although the start time of the flare is not well constrained). The flux density from the ejection decayed steadily, with re-brightening events possibly indicating internal shocks in the ejecta or interaction with ISM density enhancements. Our radio observations of J1820 track the entire evolution of the approaching ejecta, where we temporally resolve the transition from a short timescale decay phase (more typical of the timescales associated with transient soft state emission), a subsequent re-brightening, and then a long timescale decay phase (see Supplementary Information for a discussion of the decay rates, and comparison with other sources). The most likely explanation for the slowly decaying flux density is that there is constant \textit{in situ} particle acceleration as the jet decelerates via interactions with the nearby interstellar medium (ISM)\supercite{wangdai2003}. In this scenario, by the time of our energetic analysis based on the resolved emission, all of the supplied energy, $\textrm{E}_{\mathrm{f}}$, responsible for the observed radio emission would have come from this deceleration. The kinetic energy of the ejecta at a given moment is $\textrm{KE}=(\Gamma-1)\textrm{E}$, where E is the internal energy of the ejecta and $\Gamma$ is the bulk Lorentz factor. We denote the initial and final (at the time of our measurement of $\textrm{E}_{\mathrm{f}}$) internal energies and Lorentz factors by the subscripts (i,f). From the condition that deceleration has provided the observed energy, we have that $(\textrm{KE})_{\mathrm{i}}-(\textrm{KE})_{\mathrm{f}}\gtrsim\textrm{E}_{\mathrm{f}}$ or, equivalently, $(\textrm{KE})_{\mathrm{i}}=(\Gamma_{\mathrm{i}}-1)\textrm{E}_{\mathrm{i}}\gtrsim(\Gamma_{\mathrm{f}}-1)\textrm{E}_{\mathrm{f}}+\textrm{E}_{\mathrm{f}}=\Gamma_{\mathrm{f}}\textrm{E}_{\mathrm{f}}$. Given our estimates for $\textrm{E}_{\mathrm{i}}$ and $\textrm{E}_{\mathrm{f}}$, we see that $\Gamma_{\mathrm{i}}\gtrsim(\Gamma_{\mathrm{i}}-1)/\Gamma_{\mathrm{f}}\gtrsim70$. Such a large initial Lorentz factor is extremely unlikely for most jet geometries since the ejecta would be extremely Doppler de-boosted and intrinsically more luminous by orders of magnitude (in the manner of an off-axis gamma-ray burst). Therefore we must conclude that our initial estimate of the initial internal energy, $\textrm{E}_{\mathrm{i}}$, is at least two orders of magnitude too low (there is no clear way that $\textrm{E}_{\mathrm{f}}$ can have been overestimated), and that the majority of kinetic energy released is not well traced by early-time radio flaring.

Regardless of the powering mechanism, we may take $\textrm{E}_{\mathrm{f}}\sim\textrm{10}^{\mathrm{42}}\,\textrm{erg}$ as a strong lower limit to the total energy supplied to the jet, and assume that the jet was launched over a phase of $\lesssim\textrm{6.7}\,\textrm{hr}$, the rise time of the optically thin flare during the state transition. From this we derive a required energy supply rate to the launched ejection of $\textrm{4}\times\textrm{10}^{\mathrm{37}}\,\textrm{erg}\,\textrm{s}^{\mathrm{-1}}$, around 50\% of the contemporaneous X-ray luminosity.

\section*{Conclusions}

We present the following picture of the radio behaviour of J1820. A radio flare reveals the launch of relativistic transient ejecta as J1820 transitioned from the hard to soft X-ray state. We are able to track these ejecta to large separations from the black hole due to their high proper motions, and the sensitivity of MeerKAT to emission from larger angular scales. The initial fast decays (both the flare and region between MJD 58314 to MJD 58320; \prettyref{edfig:decayfits} segment one) of the approaching component is caused by the evolution of the expanding ejecta. The subsequent re-brightening (MJD 58320 to MJD 58324) and slow decay (MJD 58324 onward; \prettyref{edfig:decayfits} segment two) are the result of the ejecta continually interacting with the ISM and associated \textit{in situ} particle acceleration as jet kinetic energy is lost. The physical size of the emission $\sim\textrm{90}$ days after the ejection reveals a very large energy content in the ejecta, with an internal energy much larger than the internal energy estimated from the state transition radio flare. These observations and their interpretation present an unprecedented and comprehensive view of the life cycle of highly relativistic ejections from a stellar mass black hole over the first half a year after launch.

\clearpage

\printbibliography[segment=\therefsegment,check=onlynew]
\newrefsegment


\clearpage

\begin{figure}[H]
\centering
\includegraphics[width=0.8\columnwidth]{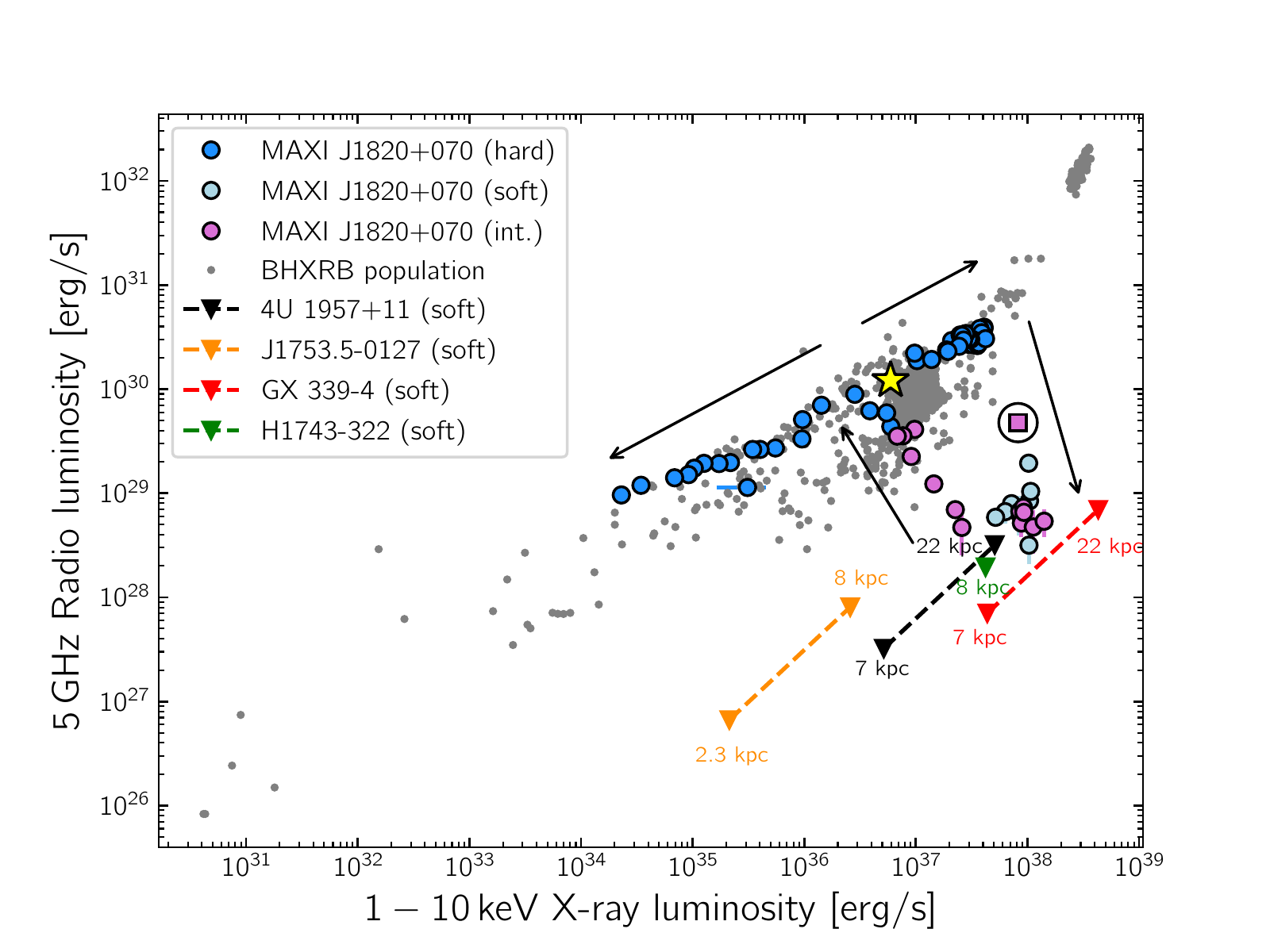}
\caption{\textbf{The radio--X-ray correlation for the BHXRB population and J1820.} Radio luminosity as a function of X-ray luminosity for J1820, based on our monitoring at $\textrm{15.5}\,\textrm{GHz}$ with the AMI-LA radio Telescope (scaled to $\textrm{5}\,\textrm{GHz}$ assuming a flat spectrum) and X-ray observations from the \textit{Swift} X-ray telescope. The data for J1820 in the hard state, soft state and intermediate state are shown by dark blue circles, light blue diamonds and purple squares, respectively. For the majority of data points the error bars are too small to be seen. The yellow star marks our first simultaneous radio/X-ray observation of J1820 ($\textrm{3.2}\,\textrm{d}$ after our first radio observation) and the black arrows show schematically the time evolution of the outburst. We only use X-ray observations within $8\,\textrm{h}$ of our radio observations, with the exception of the purple square circumscribed with a circle. In this case the observations were taken $\sim\textrm{14}\,\textrm{h}$ apart. Error bars on data points indicate one sigma uncertainties. Data from the literature on other black hole systems are indicated by grey dots\supercite{motta2018}. We mark upper limits for core soft state emission from the XRB systems 4U 1957+11, J1753.5$-$0127, GX 339$-$4 and H1743$-$322\supercite{russell2011,coriat2011,rushton2016,drappeau2017} for a range of possible distances. We do not include radio observations taken during the state transition flare. We use a distance of $\textrm{3.1}\,\textrm{kpc}$ when calculating the luminosities\supercite{atri2019}.\label{fig:lxlr}}
\end{figure}

\begin{addendum}

 \item[Correspondence] Correspondence and requests for materials should be addressed to JSB\\(joe.bright@physics.ox.ac.uk).

 \item JSB acknowledges the support of a Science and Technologies Facilities Council Studentship. ET acknowledges financial support from the UnivEarthS Labex program of Sorbonne Paris Cit{\'e} (ANR-10-LABX-0023 and ANR-11-IDEX-0005-02). DAHB acknowledges support by the National Research Foundation. PAW acknowledges support from the NRF and UCT. JCAM-J is the recipient of an Australian Research Council Future Fellowship (FT140101082), funded by the Australian government. AH acknowledges that this research was supported by a Grant from the GIF, the German-Israeli Foundation for Scientific Research and Development. IH and DRAW acknowledge support from the Oxford Hintze Centre for Astrophysical Surveys which is funded through generous support from the Hintze Family Charitable Foundation. JM acknowledges financial support from the State Agency for Research of the Spanish MCIU through the ``Center of Excellence Severo Ochoa'' award to the Instituto de Astrof\'isica de Andaluc\'ia (SEV-2017-0709) and from the grant RTI2018-096228-B-C31 (MICIU/FEDER, EU).
 
 The MeerKAT telescope is operated by the South African Radio Astronomy Observatory, which is a facility of the National Research Foundation, an agency of the Department of Science and Technology. We thank the staff of the Mullard Radio Astronomy Observatory for their invaluable assistance in the commissioning, maintenance, and operation of AMI, which is supported by the Universities of Cambridge and Oxford. We acknowledge support from the European Research Council under grant ERC-2012-StG-307215 LODESTONE. We thank the \textit{Swift} team for performing observations promptly on short notice. The National Radio Astronomy Observatory is a facility of the National Science Foundation operated under cooperative agreement by Associated Universities, Inc. e-MERLIN is a National Facility operated by the University of Manchester at Jodrell Bank Observatory on behalf of STFC. We acknowledge the use of the Inter-University Institute for Data Intensive Astronomy (IDIA) data intensive research cloud for data processing. IDIA is a South African university partnership involving the University of Cape Town, the University of Pretoria and the University of the Western Cape. We thank the International Space Science Institute in Bern, Switzerland for support and hospitality for the team meeting `Looking at the disc-jet coupling from different angles: inclination dependence of black-hole accretion observables'.
 
 \item[Author Contributions] JSB led interpretation of results, wrote a significant portion of the manuscript, and performed the reduction of the MeerKAT and AMI-LA data. RPF contributed to the interpretation of results and wrote a significant portion of the manuscript. SEM contributed to the interpretation of results, and performed the reduction of the \textit{Swift} and MAXI X-ray data. JCAM-J contributed to the interpretation of results. DW and JM performed the reduction of the eMERLIN data. RMP and JCAM-J performed the reduction of the VLA data. JCAM-J performed the reduction of the VLBA data. IH and ET assisted with the reduction of the MeerKAT observational data. DT, DG, GRS, AJT, TR and DAHB provided useful comments on the manuscript. PAW, RPA, PJG, AH, AJvdH, EGK, VAM, AR and RAMJW provided useful comments on the manuscript, and were instrumental in the organisation and implementation of the ThunderKAT large survey project. 
 
 \item[Data Availability] All radio maps used in our analysis are available from the corresponding author on reasonable request. The data used to create the radio--X-ray correlation (\prettyref{fig:lxlr}) are available in a Source Data file. The AMI-LA data of the radio flare shown in \prettyref{edfig:amiflare} are available in a Source Data file
 The authors declare that all other data supporting the findings of this study are available within the paper (and its Supplementary Information).
 
 \item[Competing Interests] The authors declare that they have no competing financial interests.
\end{addendum}

\clearpage


\clearpage

\clearpage

\begin{figure}[H]
\begin{center}
\includegraphics[width=0.8\columnwidth]{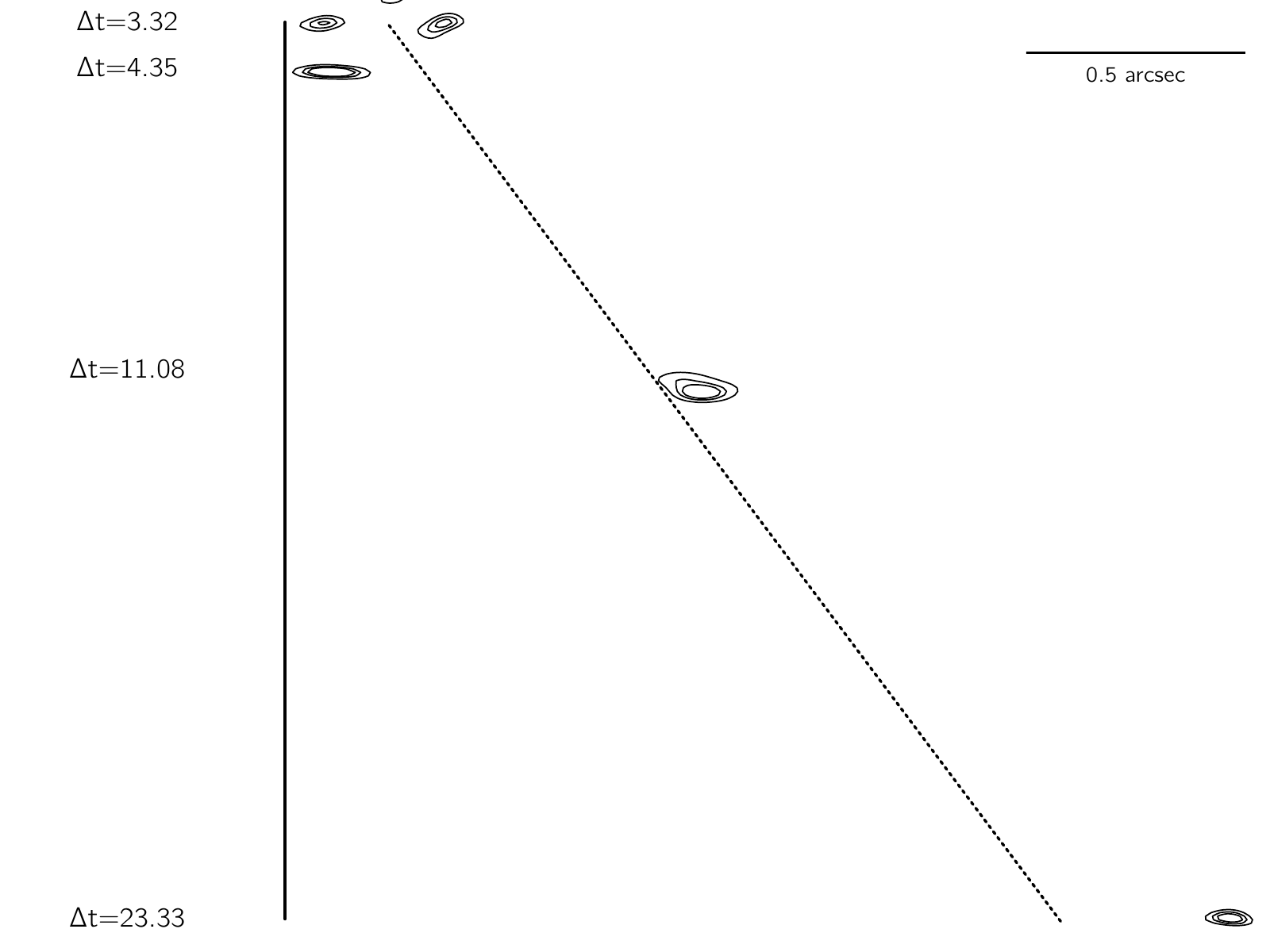}
\caption{\textbf{High angular resolution radio observations of J1820 made with eMERLIN.} eMERLIN observations of J1820 show a jet component distinct from the black hole position. The beam sizes, chronologically, are $99.2\,\textrm{mas}\times30.3\,\textrm{mas}$, $127.5\,\textrm{mas}\times27.5\,\textrm{mas}$, $106.6\,\textrm{mas}\times32.2\,\textrm{mas}$ and $130.5\,\textrm{mas}\times26.8\,\textrm{mas}$, respectively. All images have been rotated $\sim\textrm{65}^{\circ}$ anticlockwise. Contours mark ($\textrm{105}$, $\textrm{150}$, $\textrm{60}$, $\textrm{125}$) $\mu\textrm{Jy}\,\textrm{beam}^{-\mathrm{1}}\times$ $\textrm{log}(n)$ for $n = 4, 5, 6$, where the pre-factor corresponds to the images chronologically. The black vertical solid line marks the position of the core, determined from a hard state observation made with eMERLIN. The black dashed line shows the best fit ballistic trajectory of the (approaching) ejection, with the fit constrained by all observations presented in \prettyref{sitab:allpos}. $\Delta\textrm{t}$ is the time, in days, since the start of a radio flare that occurred during the hard to soft state transition ($\Delta\textrm{t}=0$ at MJD 58305.68), and is shown to the right of each observation. All observations have the same angular scale, and a scale bar is shown in the top right of the figure. Details on the data reduction procedure are presented in the Methods section, and flux densities are presented in \prettyref{sitab:allflux}.\label{fig:em}}
\end{center}
\end{figure}

\clearpage

\begin{figure}[H]
\begin{center}
\includegraphics[width=0.7\columnwidth]{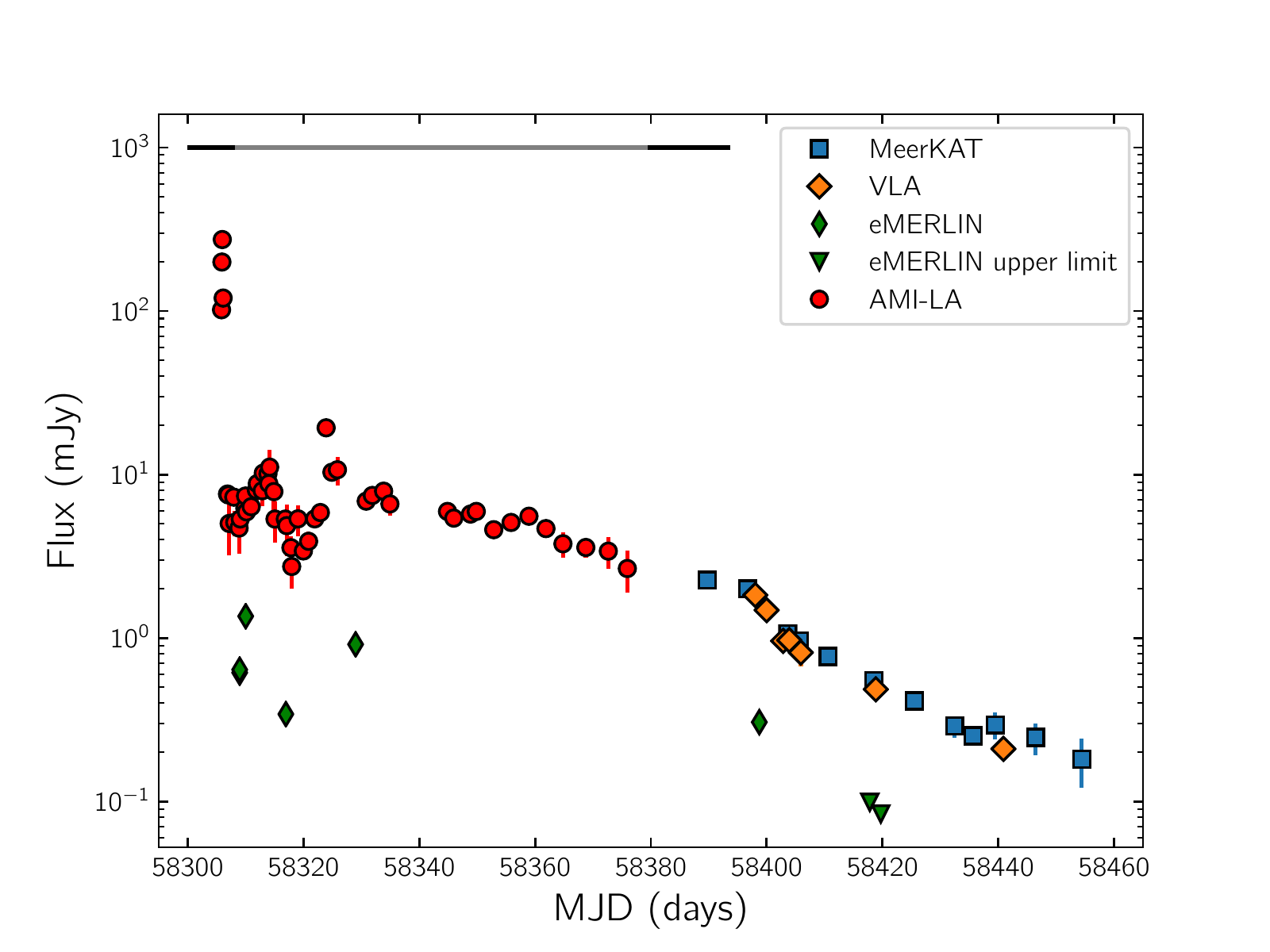}
\caption{\textbf{The radio flux density from the approaching radio ejecta over a  $\textrm{150}\,\textrm{d}$ period, starting near our inferred ejection time.} Data taken at different frequencies have been scaled by a spectral index $\alpha=-\textrm{0.7}$ ($F_{\nu}=A\nu^{\alpha}$; appropriate for optically thin synchrotron emission from jet ejecta) to a common frequency of $\textrm{1.28}\,\textrm{GHz}$. We do not scale the upper limits. The MeerKAT, eMERLIN and VLA data are measurements of the approaching jet flux density from images in which it is clearly spatially resolved from the core. We do not include AMI-LA data after J1820 returned to the hard state (around MJD 58390) as the flux density was dominated by the re-brightened core after this time. The grey horizontal line marks the duration of the soft state, and the black lines the intermediate state. Error bars on data points indicate one sigma uncertainties.\label{fig:lc}}
\end{center}
\end{figure}

\clearpage

\clearpage


\begin{methods}
\subsection{Arcminute Microkelvin Imager Large Array Observations}
We began an intensive monitoring campaign on MAXI J1820+070 (J1820) with the Arcminute Microkelvin Imager Large Array (AMI-LA\supercite{zwart2008,hickish2018}) $\textrm{1.5}\textrm{h}$ after the \textit{Swift} burst alert telescope (BAT) triggered on the source, at 00:11:39 UT on 12\textsuperscript{th} March 2018 ($t_0$ = MJD 58189.0709). The AMI-LA is robotically triggered by \textit{Swift}-BAT observations, and observes sources as soon as visibility constraints allow\supercite{staley2013}. Between $t_0$ and MJD 58462.45 we observed J1820 with the AMI-LA for a total of 183 epochs. These data were all taken at a central frequency of $\textrm{15.5}\,\textrm{GHz}$ across a $\textrm{5}\,\textrm{GHz}$ bandwidth consisting of 4096 channels, which we average down to 8 for imaging. Radio frequency interference (RFI) flagging and bandpass and phase reference calibration were performed using a custom reduction pipeline\supercite{perrott2013}. Additional flagging and imaging was performed in the Common Astronomy Software Applications (CASA\supercite{mcmullin2007}) package. For imaging we use natural weighting with a clean gain of 0.1. To measure the source flux density we use the CASA task IMFIT. The resolution of the AMI-LA (characteristic beam dimensions $40'\times30''$) when observing at the declination of J1820 mean that the source is unresolved in all epochs.

\clearpage

\subsection{Multi-Element Linked Interferometer Network}
We made observations of J1820 with the eMERLIN interferometer over the course of the 2018 outburst at $\textrm{1.5}\,\textrm{GHz}$ and $\textrm{5}\,\textrm{GHz}$, for a total of 15 epochs. Data taken at $\textrm{5}\,\textrm{GHz}$ (March \& July 2018) was done so across a $\textrm{512}\,\textrm{MHz}$ bandwidth, split into 4 spectral windows each of which consisted of 128 channels. Data taken at $\textrm{1.5}\,\textrm{GHz}$ (October) also had a $\textrm{512}\,\textrm{MHz}$ bandwidth, but instead was split into 8 spectral windows each consisting of 128 channels.

Data flagging and calibration were performed using version 0.9.24 of the eMERLIN CASA pipeline (https://github.com/e-merlin/eMERLIN\_CASA\_pipeline/) using standard calibration steps. We performed additional data flagging with AOFlagger\supercite{offringa2012}. For all of the observations, 3C286, OQ208 (QSO B1404+286) and J1813+0615 were used as the primary flux, bandpass and phase calibrators, respectively. Imaging was performed in CASA using standard procedures and natural weighting. While some observations had a bright core, allowing for the possibility of self-calibration, we opt to not perform additional calibration steps in order to preserve the absolute astrometry of our measurements. A summary of the eMERLIN observations, including participating antennas, is given in \prettyref{sitab:emerlin}.

\clearpage

\subsection{MeerKAT}
We first observed the location of J1820 with the MeerKAT radio interferometer at 17:46:40.5 UT on 28\textsuperscript{th} September 2018 (MJD 58389.74) for a total of $\textrm{15}\,\textrm{min}$ on source integration time. This observation was taken as part of the ThunderKAT large survey project\supercite{fender2017}. 62 of the 64 antennas were used in the observation, with a maximum baseline of $\textrm{7.698}\,\textrm{km}$. Data were taken at a central frequency of $\textrm{1.28}\,\textrm{GHz}$ across a $\textrm{0.86}\,\textrm{GHz}$ bandwidth consisting of 4096 channels each of width $\textrm{209}\,\textrm{kHz}$. We used J1939-6342 as the flux and bandpass calibrator. J1733-1304 was used as the complex gain calibrator, and was observed for $\textrm{2}\,\textrm{min}$ before and after the source field. Additional observations were taken with identical instrument and calibrator setups (apart from the number of antennas). Data were flagged for RFI and other issues in both the CASA and AOFlagger software packages. In CASA we flag the first and final 100 channels from the observing band, autocorrelations and zero amplitude visibilities. We then used AOFlagger to detect and remove RFI in the time and frequency domain. Flux scaling, bandpass calibration and complex gain calibration were all performed in CASA using standard procedures. After flagging and phase reference calibration the data were averaged in time ($\textrm{8}\,\textrm{s}$) and frequency (4 channels) for imaging. We used WSClean\supercite{offringa2014} to image the entire square degree field with uniform weighting, with the auto-masking threshold for deconvolution set to 4.5 times the (local) RMS flux density. We use a clean gain of 0.1. We do not perform any self-calibration on the data, despite the ample flux density in the field, in order to preserve absolute astrometry. A summary of the MeerKAT observations is given in \prettyref{sitab:meerkat}.

\clearpage

\subsection{Karl G. Jansky Very Large Array}
We observed J1820 with the VLA for a total of 10 epochs, beginning on October 7\textsuperscript{th}, after the source had returned to the hard X-ray state. All observations were taken at C band and the VLA was either in D (its most compact) or C configuration. Data reduction was performed using standard procedures (e.g. https://casaguides.nrao.edu/index.php/Main\_Page).

\clearpage

\subsection{Very Long Baseline Array}
We observed J1820 for a single epoch on MJD 58306 with the Very Long Baseline Array (VLBA). The source was observed for $\textrm{1}\,\textrm{hr}$, reaching an RMS noise of $\textrm{140}\,\mu\textrm{Jy}$. In this observation we detect both the approaching and receding jet component (the core was not detected as the source was in the soft state for this observation). Due to the high proper motion the source moves an angular distance greater than the synthesised beam of the array in one hour, and as such is `smeared' along the direction of motion. To measure the positions of the two components we fit the main peak of the flux profile along the jet axis with a Gaussian, using the centroid as the position. To estimate the error on the position, we smooth the entire flux profile of each ejection along the jet axis using a Savitzky-Golay filter until the profile is Gaussian-like. We then fit this smoothed profile with a Gaussian, using the half width at half maximum as the error. These measurements are reported in \prettyref{sitab:allpos}.

\clearpage

\subsection{Radio Positions}
A critical part of our analysis relies on measuring the positions of the core and ejections from J1820 with a range of telescopes. For our observations with eMERLIN, MeerKAT and the VLA in C configuration we fit the sources in the image plane using the CASA task IMFIT. For MeerKAT observations we attempt to fit three point source (fixed beam major and minor axes and position angle) components, allowing the position and amplitude to vary. For MeerKAT observations where a three component fit would not converge (early time observations when the receding jet had a small angular separation), we fit two components instead. We do not fix the core position in our MeerKAT analysis to the known position from our eMERLIN observations, so any systematic position errors will affect all components and be negated when calculating the separation. We used the same procedure for the VLA C configuration data. For eMERLIN observations the components are separated significantly and as such can be boxed and fit individually using IMFIT. When fitting the ejection components we do not fix the dimensions of the elliptical Gaussian used by IMFIT, as the ejection components are not point-like. We do fix the size of the component used to fit the (known to be compact) core. As core emission was not detected in all eMERLIN observations (due to core quenching in the soft state) we use the position measured from a bright observation on MJD 58201 to calculate the separation. We did not use the position errors reported by IMFIT for analysis, as we found these tended (especially for bright components) to be many times smaller than the synthesised beam. While the centroid of an elliptical Gaussian is known to an accuracy determined by the ratio of the synthesised beam dimensions to the signal to noise ratio of the Gaussian, this is only true to a certain accuracy level before absolute astrometric uncertainties begin to dominate. For example, it is recommended for the VLA (https://science.nrao.edu/facilities/vla/docs/manuals/oss/performance/positional-accuracy) that, unless special calibration steps are taken, positions are not reported to an accuracy of more than about 10\% of the synthesised beam width. For all of our observations we report position errors as $A/\sigma$, where A is the amplitude of the fit component and $\sigma$ is the width of the synthesised beam at an angle connecting the fitted component with its corresponding ejection/core component, but never to an accuracy greater than 10\% of this width (we confirm using check sources in the MeerKAT field that the position errors calculated as such are sensible). There are two exceptions to this. When, in our eMERLIN observations, only the core was detected we simply report the IMFIT RA and Dec. errors, combined in quadrature. For eMERLIN observations when only an ejection component was detected we use the observation taken on MJD 58201 for the purpose of finding the angle at which to calculate $\sigma$. For VLA observations taken when the array was in the more compact D configuration the resolution was not good enough to fit sources in the image plane. For these observation we performed fitting in the UV plane using the CASA task UVMULTIFIT\supercite{martividal2014}, after building a sky model and subtracting background sources using the CASA task UVSUB. When fitting UV plane components we fix the spectral index of the ejecta to be --0.7, but allow the core spectral index to vary as a free parameter. Components were all specified to be point sources. The results of the positions and flux densities measured from this analysis are presented in \prettyref{sitab:allpos} and \prettyref{sitab:allflux}, respectively. The position errors from our VLBA observation are described in a separate Methods sub-section. We do not correct the eMERLIN observations for the proper motion of the core as the change in separation caused by this motion ($\lesssim3\,\textrm{mas}$) is $\lesssim1\%$ of the separation for all epochs, and is significantly less than the eMERLIN separation errors which are $\geq15\,\textrm{mas}$.

\clearpage

\subsection{\textit{Swift}-XRT}
J1820 was observed at high brightness levels during a large fraction of its outburst phase with the \textit{Swift} X-ray Telescope (XRT\supercite{gehrels2004}). For this reason most of the observations considered in this work have been taken using the XRT Window Timing (WT) mode, which provides one-dimensional imaging with a $\textrm{1.7}\,\textrm{ms}$ time resolution, and allows bright sources to be observed. Photon pile-up is known to induce distortion of XRT's spectral response (see http://www.swift.ac.uk/\\analysis/xrt/pileup.php) and it starts to have non-negligible effects at a nominal count rate threshold of approximately $\textrm{150}\,\textrm{counts}\,\textrm{s}^{\mathrm{-1}}$. Since J1820 was most of the time observed with count rates in XRT significantly higher than this threshold, \textit{Swift}/XRT data are often significantly affected by photon pile-up. Therefore, following the recommendation found in the \textit{Swift}/XRT data reduction threads (http://www.swift.ac.uk/analysis/xrt/\#abs), we extracted only grade 0 events from the raw data. This helps to mitigate the effects of pile-up in bright sources (http://www.swift.ac.uk/xrt\_\\curves/cppdocs.php) and reduces the spectral distortion encountered in WT mode below $\textrm{1.0}\,\textrm{keV}$ in case the energy spectrum is highly absorbed. Furthermore, we ignored data below $\textrm{0.6}\,\textrm{keV}$ as below this energy the energy spectra can be dominated by strong redistribution effects associated with the WT readout process, and by trailing charge released from deep charge traps in the CCD on time-scales comparable to the WT readout time, which results in additional (spurious) low-energy events.

In order to exclude the regions of the detector where pixels were pile-up saturated, we extracted events in circular regions centred at the source position, with variable inner radius and outer radius fixed to 20 pixels. We determined the final inner radius of the extraction region by varying it until the spectral shape was no longer changing as a function of the inner radius itself, and the count rate was lower than approximately $\textrm{150}\,\textrm{counts}\,\textrm{s}^{\mathrm{-1}}$.

The spectral evolution of J1820 was generally slow (save for the times of the hard-to-soft state transition, which was missed by \textit{Swift}), hence we extracted one spectrum per \textit{Swift}/XRT observation. We obtained a total of 80 spectra for X-ray observations within $\textrm{8}\,\textrm{h}$ of our AMI-LA observations, covering the entire duration of the outburst. Since in this work we are interested in the source luminosity rather than its detailed spectral properties (to be presented in a future work), we fitted each spectrum between $\textrm{0.6}\,\textrm{keV}$ and $\textrm{10}\,\textrm{keV}$ using the XSPEC package\supercite{arnaud1996} with a phenomenological model constituted by a simple power law model combined with a multi-colour disc-blackbody (power law + disk in XSPEC), both modified by interstellar absorption (TBNEW\_FEO). We then measured the source flux in the $\textrm{1}$-$\textrm{10}\,\textrm{keV}$ energy band, which we converted the flux to luminosity using a distance from the source of $\textrm{3.8}\,\textrm{kpc}$.

\end{methods}

\clearpage

\printbibliography[segment=\therefsegment,check=onlynew]
\newrefsegment

\clearpage


\begin{extdatafig}[H]
\begin{center}
\includegraphics[width=0.7\columnwidth]{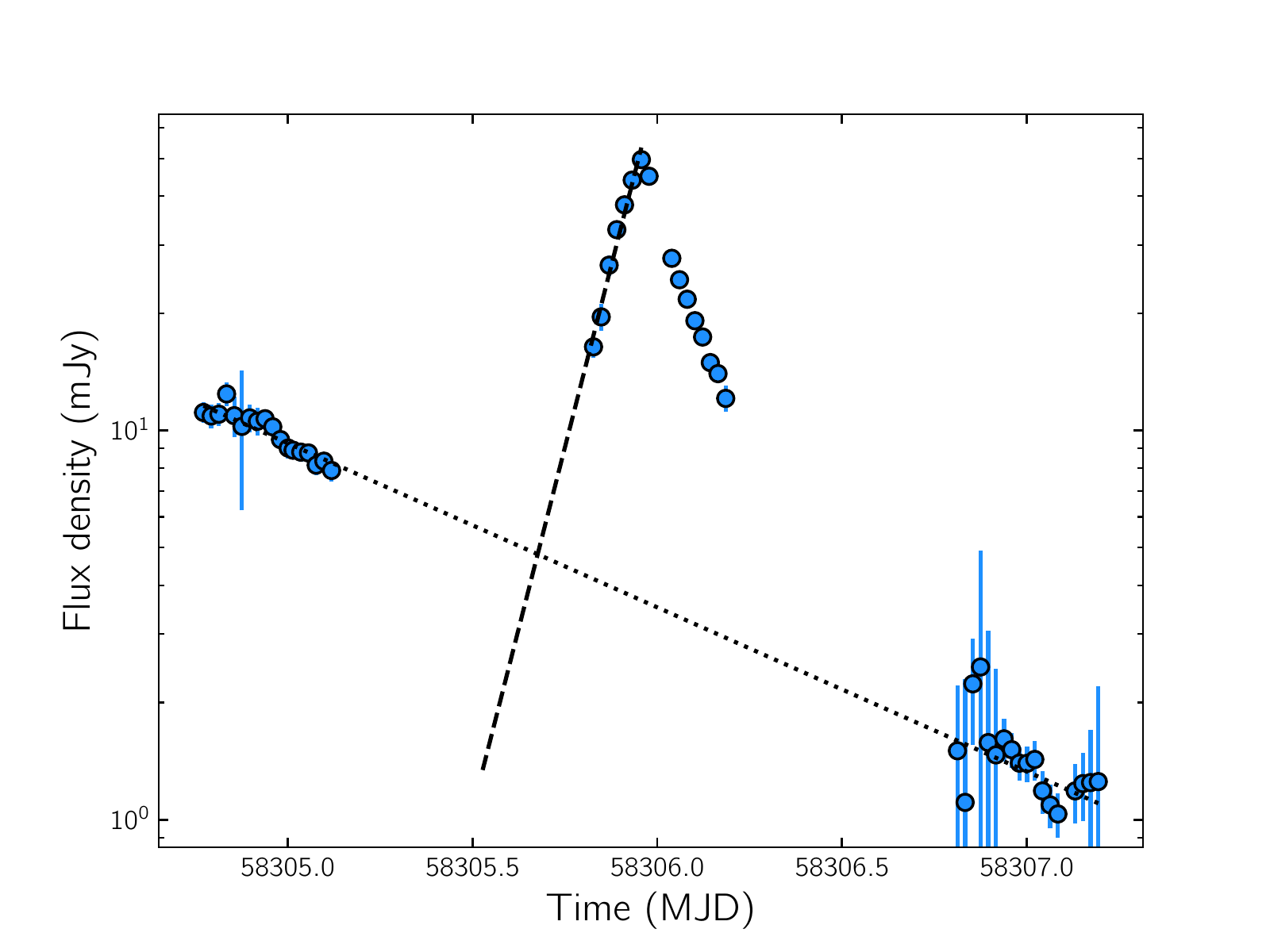}
\caption{\textbf{AMI-LA observations of a state transtion radio flare from J1820.} AMI-LA observations of a radio flare which occurred as J1820 transitioned from the hard to soft X-ray state. The blue data points correspond to $\textrm{30}\,\textrm{min}$ of (\textit{u},\textit{v}) amplitudes averaged over all baselines and frequencies. The errors on individual points include a statistical error (calculated from the standard deviation of data within the $\textrm{30}\,\textrm{min}$ bin) and a 5\% calibration uncertainty, combined in quadrature. Dotted and dashed lines show exponential fits to the core quenching and the rise of the flare, respectively. We use these to estimate the rise time of the flare, which we take as the time between the intercept of these fits and the peak data point of the flare, as well as its start date. Error bars on data points indicate one sigma uncertainties.\label{edfig:amiflare}}
\end{center}
\end{extdatafig}

\clearpage

\begin{extdatafig}[H]
\begin{center}
\includegraphics[width=0.7\columnwidth]{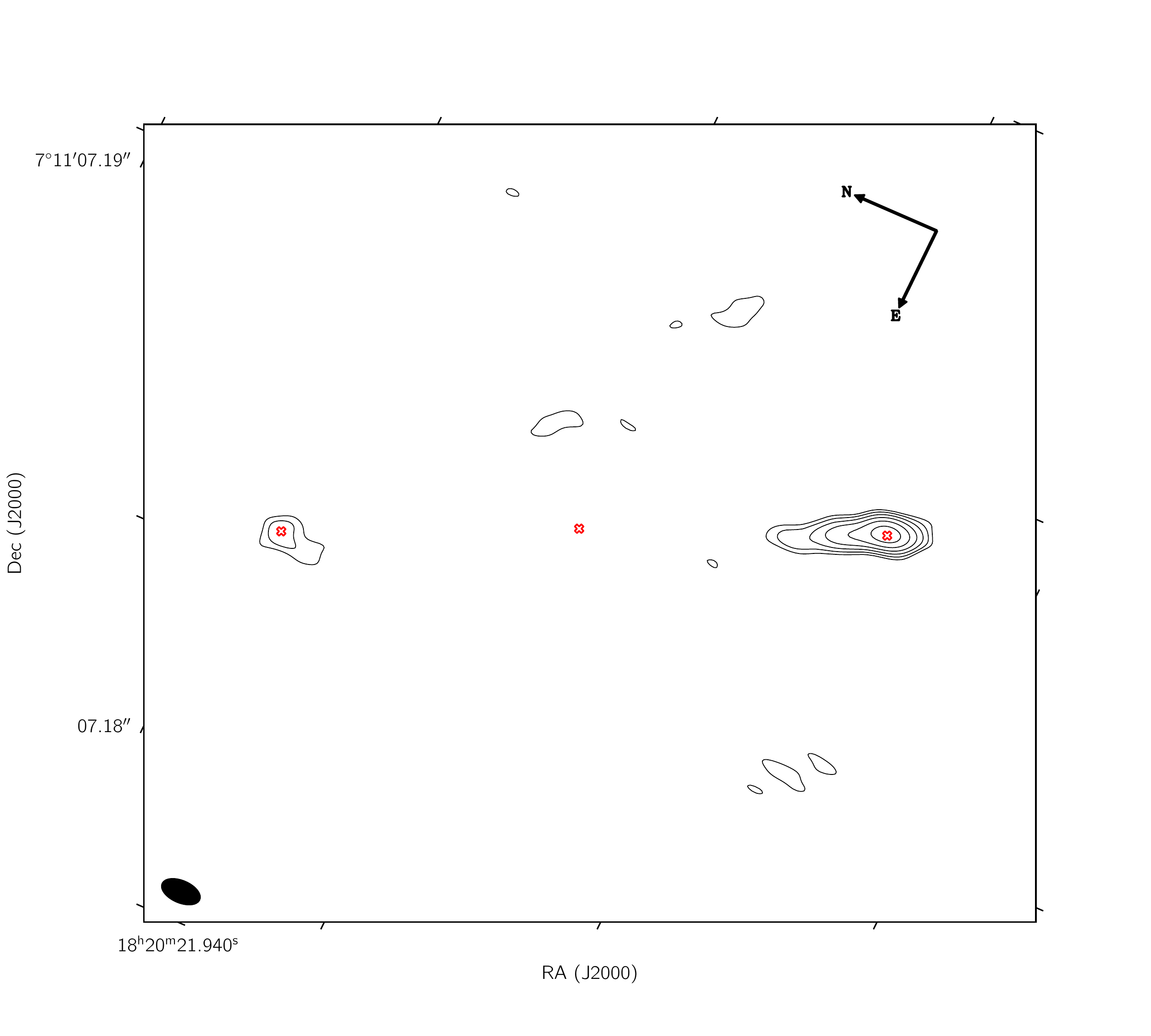}
\caption{\textbf{A VLBA observation of J1820 from MJD 58306.22.} Contours mark $\textrm{140}\,\mu\textrm{Jy}\times(\sqrt{\textrm{2}})^{n}$ for $n=3,4,5,6,7,8,9$. We mark the position of the core (central red cross; inferred from previous hard state observations) and the measured positions of the approaching (red cross to the right of the core) and receding (red cross to the left of the core) jet from the image. These are given in Table \prettyref{sitab:allpos}. The black ellipse in the bottom left corner shows the synthesised beam with a major and minor axis of 0.0009$''$ and 0.0005$''$, respectively.\label{edfig:vlbi}}
\end{center}
\end{extdatafig}

\clearpage

\begin{extdatafig}[H]
\begin{center}
\includegraphics[width=0.7\columnwidth]{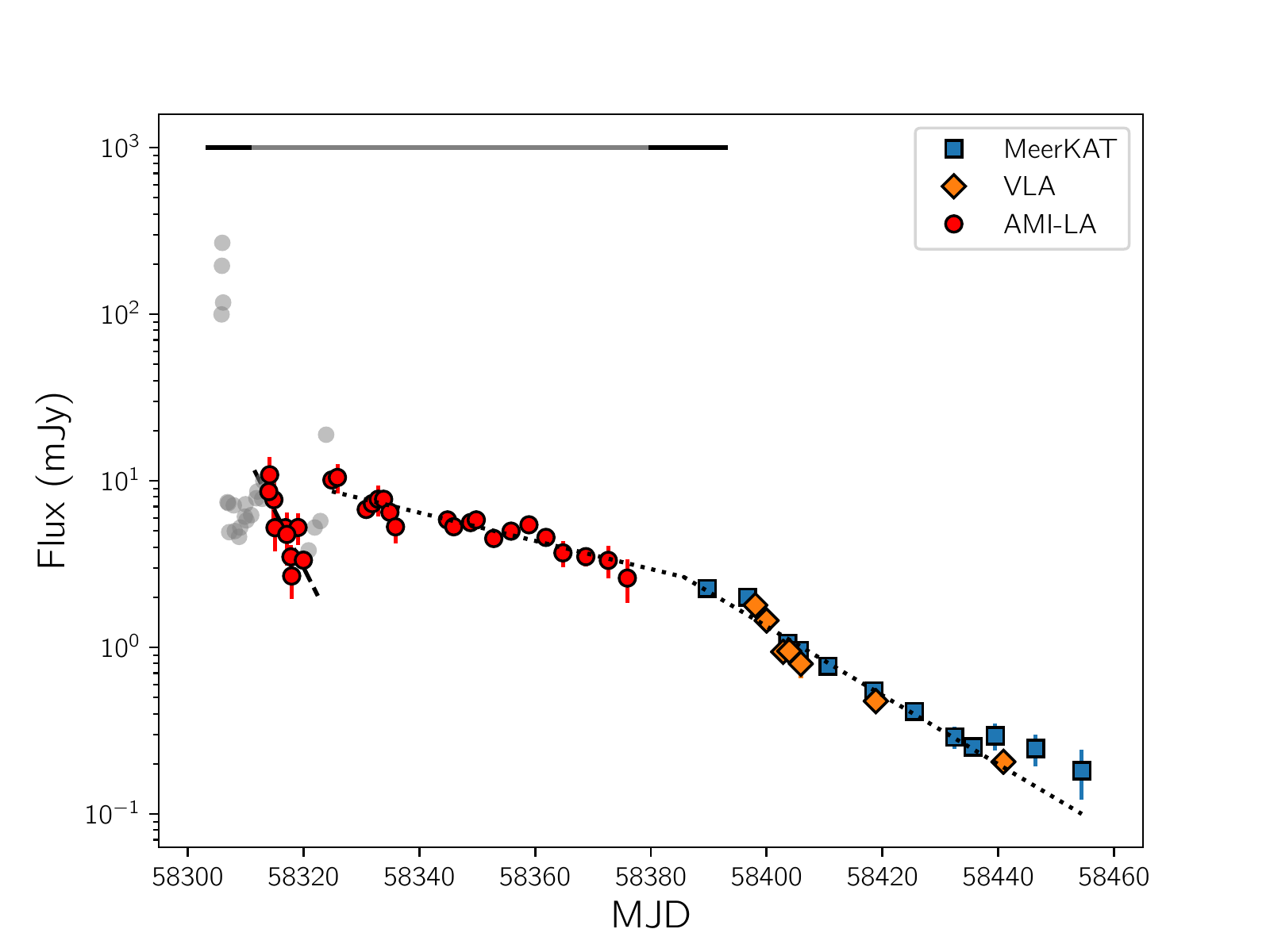}
\caption{\textbf{The radio flux density from the approaching radio ejecta over a $\textrm{150}\,\textrm{d}$ period, starting near our inferred ejection time.} As with \prettyref{fig:lc}, with the eMERLIN and VLBA data removed. We fit sections of the light curve with exponential decay functions of the form $F_{\nu}=Ae^{-\Delta t/\tau}$. Data shaded grey are not included in the fitting. The first light curve segment (fast decaying AMI-LA data; MJD 58314 to 58320), is well described ($\chi^{2}_{\nu}=\textrm{1.21}$) by a decay with a characteristic time scale of $\textrm{6}\pm\textrm{1}\textrm{d}$ (dashed line). We opt to fit the apparently slower decay (MJD 58324 onward) with a broken exponential function (dotted line). The best fit decay rates are $\textrm{51}\pm\textrm{6}\,\textrm{d}$ and $\textrm{21.0}\pm\textrm{0.9}\,\textrm{d}$, with the break occurring at MJD $\textrm{58386}\pm\textrm{4}$ ($\chi^{2}_{\nu}=\textrm{1.59}$). Error bars on data points indicate one sigma uncertainties.\label{edfig:decayfits}}
\end{center}
\end{extdatafig}

\clearpage

\begin{extdatafig}[H]
\begin{center}
\includegraphics[width=0.7\columnwidth]{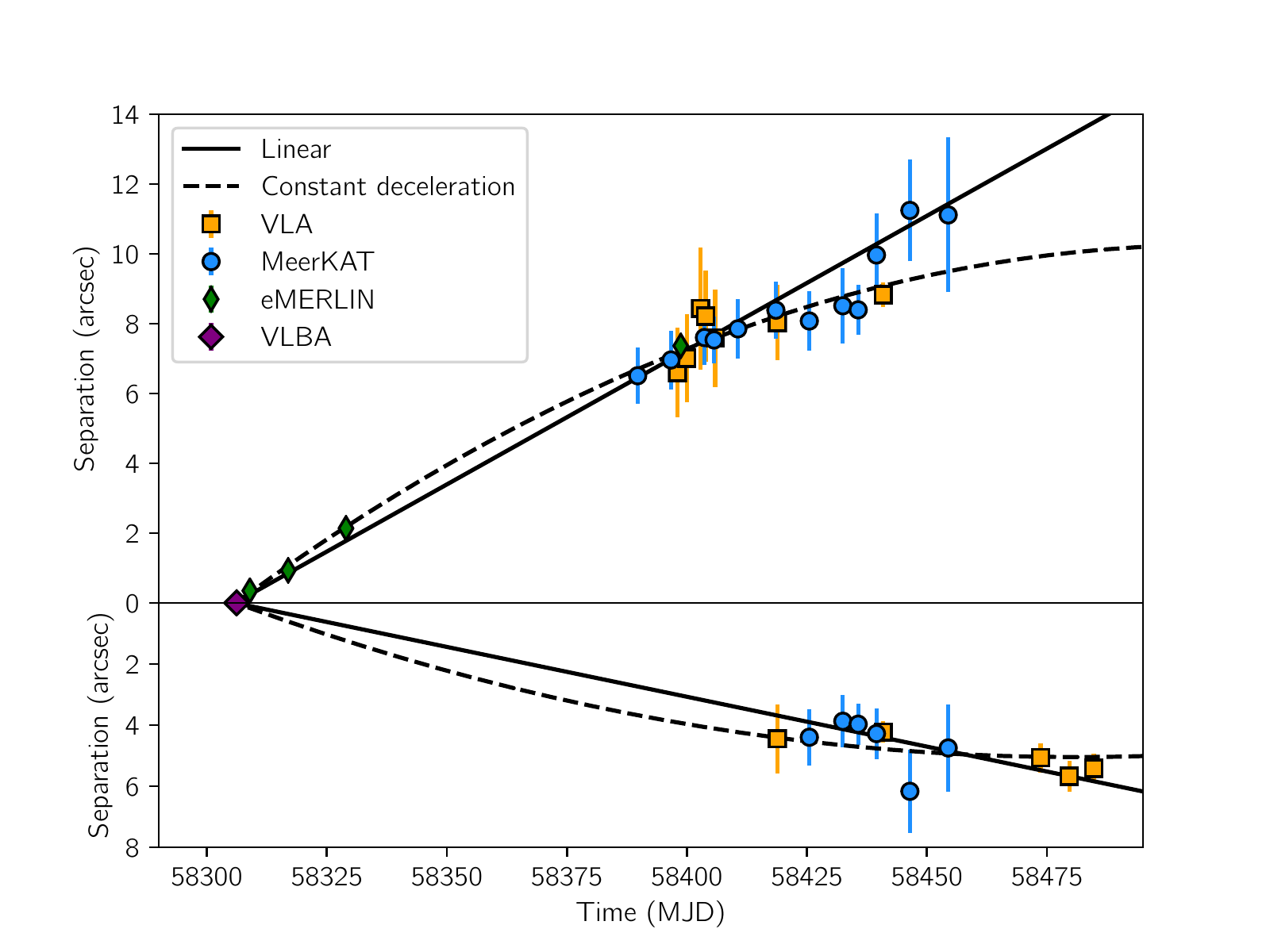}
\caption{\textbf{The angular separation evolution of the approaching and receding jet components.} The angular separation of the approaching (top panel) and receding (bottom panel) ejections from J1820 with time. We jointly fit both the approaching and receding jet motion with two models. Firstly we assume that both components propagate with ballistic motion and were launched simultaneously. For this case we find $\mu_{\textrm{app}}=\textrm{77}\pm\textrm{1}\,\textrm{mas}\,\textrm{d}^{\mathrm{-1}}$, $\mu_{\textrm{rec}}=\textrm{33}\pm\textrm{1}\,\textrm{mas}\,\textrm{d}^{\mathrm{-1}}$ and $t_{\textrm{launch}}=\textrm{58305.89}\pm\textrm{0.02}$ ($\Delta\textrm{t}=\textrm{0.21}\pm\textrm{0.02}$) (quantities correspond to the approaching jet velocity, the receding jet velocity and the launch time, respectively). The best fit for this model are shown by the solid black lines in the top and bottom panel. Assuming now as above, but allowing for the proper motion of each component to undergo constant deceleration, we find $\mu_{\textrm{app,0}}=\textrm{101}\pm\textrm{3}\,\textrm{mas}\,\textrm{d}^{\mathrm{-1}}$, $\mu_{\textrm{rec,0}}=\textrm{58}\pm\textrm{6}\,\textrm{mas}\,\textrm{d}^{\mathrm{-1}}$, $t_{\textrm{launch}}=\textrm{58306.03}\pm\textrm{0.02}$ ($\Delta\textrm{t}=\textrm{0.35}\pm\textrm{0.02}$), $\dot{\mu}_{\textrm{app}}=-\textrm{0.49}\pm\textrm{0.06}\,\textrm{mas}\,\textrm{d}^{\mathrm{-2}}$ and $\dot{\mu}_{\textrm{rec}}=-\textrm{0.33}\pm\textrm{0.07}\,\textrm{mas}\,\textrm{d}^{\mathrm{-2}}$ (quantities correspond to the initial approaching jet velocity, the initial receding jet velocity, the launch time, the deceleration of the approaching jet and the deceleration of the receding jet, respectively). Error bars on data points indicate one sigma uncertainties.\label{edfig:propermotion}}
\end{center}
\end{extdatafig}


\clearpage
\section*{Supplementary Information}
\subsection{Radio flare}

As MAXI J1820+070 (J1820) transitioned from the hard to soft X-ray states (i.e. was in the intermediate state) we observed a flaring event with the AMI-LA which lasted for $\sim\textrm{12}\,\textrm{h}$. In \prettyref{edfig:amiflare} we demonstrate a subsection of our AMI-LA observations which covered the flare. In order to temporally resolve the event (which occurred over the typical timescale of a single observing track with the AMI-LA) we plot the amplitude of the (\textit{u},\textit{v}) data directly, averaging over baselines and spectral windows in $\textrm{30}\,\textrm{min}$ time bins. Modelling the radio emission as a flare (caused by a discrete relativistic ejection) superposed on a constantly decaying component (due to the compact core jet quenching) we estimate the amplitude and the rise time of the flare to be $\sim\textrm{46}\,\textrm{mJy}$ and $\textrm{6.7}\pm\textrm{0.2}\,\textrm{h}$, respectively. Our estimated time that the flare began is MJD $\textrm{58305.68}\pm\textrm{0.01}$. When referring to observations of the bi-polar ejections, the observation time is taken with respect to the start of this flare. We note, however, that flares that peak due to optical depth effects are known to rise quicker, and peak first, at higher frequencies and so sub-mm observations of radio flares are likely a better proxy of the launch date of relativistic ejection\supercite{tetarenko2017}. Is is also possible that the flares are result of internal shocks, and there is a delay between the launch of ejections and collisions\supercite{fender2009}.

\subsection{Soft state decay rates}
While BHXRBs are in the soft accretion state the compact core jet is significantly quenched and any radio emission from it drops by many orders of magnitude (always below observing sensitivity limits) or switches off completely. Radio emission observed during the soft state is almost certainly associated with ejections launched during the hard to soft state transition. This radio emission is transient, and is seen to fade as the ejections expand and cool. The e-folding decay timescale (which we will hereafter refer to as simply the decay timescale) of the emission from the ejections is seen to vary significantly between sources, but can broadly be categorised as either short (decay timescales from a few to $\sim10\,\textrm{d}$) or long (decay timescales from a few tens to hundreds of days). Short decay timescales are thought to be the result of ejecta expanding and cooling, with minimal ongoing energy injection resulting from interactions from the ISM. When longer decay rates are seen, it is thought that ongoing ISM interaction provides a source of particle acceleration, partially offsetting cooling, and results in the slowed flux decline. Example of fast decays include GRS 1915 during its 1994 and 1997 outbursts, showing decay timescales of $\sim\textrm{7}\,\textrm{d}$ and $\sim\textrm{2}\,\textrm{d}$, respectively\supercite{mirabel1994,fender1999}. XTE J1748$-$288 showed a radio flux density decay timescale of $\sim\textrm{7}\,\textrm{d}$ at the start of the soft state during its 1998 outburst\supercite{brocksopp2007}. Slow decays have been seen in XTE J1550$-$564, which showed a flux density decay timescale of $\sim\textrm{85}\,\textrm{d}$ (at $1.4\,\textrm{GHz}$) following a plateau period\supercite{migliori2017}. This decay rate appeared to be wavelength dependent, with X-ray observations revealing an exponential decay rate of $\sim\textrm{340}\,\textrm{d}$ from the same ejection component. An ejection from H1743$-$322 decayed with a timescale of $\lesssim\textrm{28}\,\textrm{d}$.\\

In the main text we discuss the decay rate of the approaching ejection from J1820, which was seen to evolve throughout the soft state. To demonstrate the different decay rates we present a modified version of \prettyref{fig:lc} (\prettyref{edfig:decayfits}) from the main text, in which we fit for the decay timescale for different segments of the light curve. The first segment, between MJD 58314 and MJD 58320, shows a decay timescale of $\textrm{6}\pm\textrm{1}\textrm{d}$. In the main text we refer to this as a `fast decay'. The second segment, dates after MJD 58320, we fit with a broken exponential which shows decay rates of $\textrm{51}\pm\textrm{6}\,\textrm{d}$ and $\textrm{21.0}\pm\textrm{0.9}\,\textrm{d}$ with a break occurring at MJD $\textrm{58386}\pm\textrm{4}$. In the main text we refer to these as a `slow decay'. Both of these decay phases are also seen in our eMERLIN data, although due to the course sampling we do not include them in \prettyref{edfig:decayfits}. It appears, therefore, that the approaching ejection from J1820 initially showed a period of unimpeded cooling, followed by a long and slow decay caused by continued ISM interaction. We note that for both XTE J1550--564 and H1743--322 the decay rate of the ejecta were frequency dependent, with higher frequencies decaying slower\supercite{migliori2017,corbel2005}. This is similar to what we see for J1820, with the slower decay rate corresponding to the higher frequency (AMI-LA) data. The short delay between the ejection launch and this slow decay phase (in contrast to XTE J1550-564) may indicate that J1820 is not contained within an ISM cavity (and the decay is due to ongoing ISM interaction from the outset), or, if present, such a cavity may have a significantly smaller radius causing an earlier transition to the slow decay phase. The cause of the rise in flux between the two light curve segments (and between the end of the flare and the start of the first segment) is uncertain, but could be indicative of multiple ISM density enhancements.

The measured time of the break in the second light curve segment is remarkably close to the date where J1820 returned to the hard X-ray state (MJD 58393), and the core jet turned back on. For the two events to be connected there would have to be transport of information between the core and the approaching jet (separated by $\sim\textrm{7}''$ at this epoch) on a $\sim\textrm{7}\,\textrm{d}$ timescale. This would require an extremely high inferred proper motion of $\sim\textrm{1}\,\textrm{arcsec}\,\textrm{day}^{\mathrm{-1}}$ ($\textrm{22}\,c$ at $\textrm{3.8}\,\textrm{kpc}$). This is obviously significantly superluminal, and we would require the approaching ejection component to have a small angle (maximum $\theta\sim\textrm{5}^{\circ}$) to the line of sight for the actual velocity to be at or less than c. This angle is not compatible with the one that we measure from our fitted proper motions. It is more likely that the difference in decays is either due to the fact that the AMI-LA is probing much larger angular scales, or that contamination from the receding jet (which is contained within the AMI-LA synthesised beam) is altering the decay rate. While we have no direct measurement of the flux density from the receding jet during the AMI-LA observations presented in \prettyref{fig:lc} and \prettyref{edfig:decayfits}, we note that the receding jet is not detected in any of our eMERLIN observations and is below $\sim\textrm{600}\,\mu\textrm{Jy}$ in our MeerKAT observations and so is likely to be a significantly less dominant component.

\clearpage

\subsection{Proper Motions}
In \prettyref{sitab:allpos} we present the measured positions for the core, approaching ejection and receding ejection that we use to fit for the proper motions of each ejection (for details on this procedure see the Methods section). The angular separation with time for the two ejections is presented in \prettyref{edfig:propermotion}.

We opt to exclude measurements made from two of our eMERLIN epochs. These are marked in \prettyref{sitab:allpos}, and correspond to the smallest angular separation component in the first observation demonstrated in \prettyref{fig:em} and the second observation shown in \prettyref{fig:em}. Between these two observations this component moves with a proper motion of $\sim\textrm{30}\,\textrm{mas}\,\textrm{d}^{\mathrm{-1}}$, and was therefore launched around the same time as the faster approaching ejecta. It is evident, however, that this component is not well described alongside the rest of our measurements for either a linear or decelerating fit. Due to our lack of observations at multiple angular resolutions at this epoch, we cannot be sure if the two components detected in our first eMERLIN observation are part of a larger structure, the details of which we resolve out, or if they are distinct ejections. It is possible (though unlikely) that we missed a flare (and potentially associated ejection) with our AMI-LA monitoring, or that a single flare actually corresponded to a complex ejection morphology\supercite{tetarenko2017}. In this case the early time eMERLIN observations could be probing this morphology, and the later time data reveals the motion of the aggregated structure. We note that we could use the smaller angular separation component in our initial eMERLIN observation (MJD 58308; $\Delta t=\textrm{3.32}\,\textrm{d}$), instead of the larger angular separation component. While this provides a better fit to the first three eMERLIN observations (not underestimating the position of the component observed on MJD 58329; $\Delta t=\textrm{23.33}\,\textrm{d}$) it requires a significant deceleration to fit the entire data set. The inclusion of deceleration is not in itself an issue, however including this component when fitting both a linear and decelerating fit provide a launch date significantly \textit{after} the radio flare observed by AMI. Additionally, the observation on MJD 58310 ($\Delta t=\textrm{4.35}\,\textrm{d}$) shows a component that is consistent with the smaller angular separation component on MJD 58308 as discussed above. Finally, our VLBA observation made earlier than our eMERLIN observation on MJD 58306 reveal a component is already present, well before this inferred launch date.

It is important to attempt to account for systematic uncertainties that arise when measuring the positions of components observed at very different angular scales. There is no guarantee that the centroid of the emitting region is the same on these different angular scales when a significant amount of the flux density is resolved out, as is the case for the approaching ejection component here (the receding component was only measured quasi-simultaneously by telescopes with similar angular resolutions). Using the ratio of beam size to signal to noise for the positional error will cause the eMERLIN data to be artificially over constraining given the previous argument, so instead we derive errors based on physics considerations. Considering the ejection as a spherical region expanding at a speed of $\sim\textrm{0.05}\,c$, launched at the start of the flare observed with the AMI-LA during the hard to soft state transition, we estimate the emitting region would have an angular size of $\textrm{0.015}''$,  $\textrm{0.051}''$, $\textrm{0.11}''$, and $\textrm{0.42}''$ on MJD $\textrm{58308.98}$, $\textrm{58316.96}$, $\textrm{58329.00}$ and $\textrm{58398.73}$, respectively, and use these values as our separation error. For the final observation we cap the error at $\textrm{0.2}''$ as it is now comparable with the position error derived from our lower resolution images.

We have demonstrated the results of fitting the angular separation with both a linear proper motion model, and one with constant deceleration. Determining the statistically appropriate model for data with vastly different error bars is challenging. Even when reevaluating the errors on our eMERLIN measurements, the error on the position for these observations (especially the ones only a few weeks after the launch of the approaching ejection) are significantly smaller than those made with the VLA and MeerKAT. This is also true for the VLBA observation. Adding a free parameter to our proper motion model (e.g. a deceleration) will essentially serve only to fit the early-time eMERLIN/VLBA observations, with other data barely constraining the model. There is also the issue that the centroids of the emitting regions do not necessarily align on the very different angular scales, and as such any inferred deceleration is not necessarily the physical deceleration of the ejections. It is also worth noting that different proper motions have been reported for the jets in XRB GRS 1915+105 from observations taken with different angular resolution, and do not necessarily imply that deceleration is occurring\supercite{fender1999,rodriguez1999}. We consider both models in the text, but note that the parallax distance\supercite{atri2019} for J1820 is strong evidence against the deceleration model being required to fit this data set.

\clearpage
\subsection{Radio -- X-ray Correlation}
Quasi-simultaneous X-ray and radio observations of accreting black hole X-ray binaries have been used to establish a connection between the accretion process and the production of jets, particularly the continuously-replenished relativistic jets typically observed in the hard states (and in quiescence). Particularly well-known is the non-linear correlation between the X-ray and the radio luminosity, originally discovered in the black hole X-ray binary GX 339$-$4\supercite{corbel2003}. This correlation was initially considered universal\supercite{gallo2003}, however, more recently it has become clear\supercite{gallo2012,gallo2014} that some BHXRBs are considerably less luminous in the radio band than the canonical sources such as GX 339$-$4, and populate a second track in the radio--X-ray plane that is known as the radio-quiet track, as opposed to the original track, which is referred to as the radio-loud track. While some sources (e.g. H1743$-$322\supercite{coriat2011}) clearly follow an alternative track, it is not unambiguous that the whole population of BHXRBs can be separated into a bi-modal distribution of tracks\supercite{gallo2012}.

Attempts to identify the physical origin of the existence of such tracks have been so far unsuccessful. Differences have been explained in terms of different jet magnetic field configurations\supercite{casella2009}, the accretion flow radiative efficiency\supercite{coriat2011} or in the contribution from an additional inner accretion disc\supercite{meyerhofmeister2014}. More recently, it was proposed that the morphology of the distribution is the result of an inclination effect, which, however, remains to be confirmed by more observations of black hole X-ray binaries in the hard state\supercite{motta2018}, although we note that J1820 goes against the proposed trend. 

During the initial hard-state, J1820 travelled along the radio-loud track following a power law of the form $L_R = AL_{X}^{\alpha}$, with $\alpha=\textrm{0.42}\pm\textrm{0.05}$. The correlation showed the same slope throughout the long initial hard state, all the way up to X-ray and radio luminosities of $\sim\textrm{4}\times\textrm{10}^{\mathrm{37}}\,\textrm{erg}\,\textrm{s}^{\mathrm{-1}}$ and $\sim\textrm{6}\times\textrm{10}^{\mathrm{30}}\,\textrm{erg}\,\textrm{s}^{\mathrm{-1}}$, respectively. During the intermediate state J1820 left the radio loud track, with its radio emission dropping rapidly. The source was then detected continually throughout the soft state (although we determine this does not represent a connection between accretion and core-jet emission). We then track the core-jet turning back on as J1820 returns to the radio loud correlation, following a track with $L_R = AL_{X}^{\mathrm{-1.4}\pm\mathrm{0.4}}$, and joining at a similar location to our first quasi-simultaneous radio/X-ray detection. The radio--X-ray correlation during the end-of-outburst hard state shows $\alpha=\textrm{0.37}\pm\textrm{0.03}$, consistent with (but slightly shallower than) that on the initial hard state. A joint fit of the initial and final hard state radio--X-ray correlation returns a slope of $\alpha=\textrm{0.50}\pm\textrm{0.09}$.

Our simultaneous radio and X-ray monitoring ended on MJD 58439 at which point we measure, with the VLA, the receding jet flux density to be around 20\% of the core flux density at 6 GHz. Assuming the core has a flat spectrum\supercite{broderick2018} and the ejection is optically thin with a spectral index of --0.7, we estimate that the ejection could be contributing around 10\% of the flux density measured by the AMI-LA by this date. Fifteen days previous, a detection of the core and receding ejection with MeerKAT at 1.28 GHz measured the receding component flux density to be around 30\% of the core flux density. Under the same assumptions this would imply around a 5\% contribution to the AMI-LA flux density at this epoch. Removing (quasi-)simultaneous observations after MJD 58424 alters the slope during to second hard state to $L_R = AL_{X}^{\mathrm{0.34}\pm\mathrm{0.06}}$, and the jointly fit slope becomes $L_R = AL_{X}^{\mathrm{0.55}\pm\mathrm{0.02}}$. We conclude that the slopes are not being significantly altered by the presence of ejecta components contaminating the AMI-LA measurements of the core.

\clearpage

\printbibliography[segment=\therefsegment,check=onlynew]

\clearpage

\begin{ThreePartTable}
\tiny
\begin{longtable}{ccccccccccc}
\caption{\textbf{Positions of the core, approaching ejection, and receding ejection from the 2018 outburst of J1820.} Positions of the approaching jet, receding jet, and core components for observations with eMERLIN, MeerKAT, the VLA and the VLBA. Dates report the observation mid-point.\label{sitab:allpos}}
\\
\hline
& \multicolumn{3}{c}{Core} &  \multicolumn{3}{c}{Approaching ejection} & \multicolumn{3}{c}{Receding ejection} & \\
Date & RA & Dec & Error & RA & Dec & Error & RA & Dec & Error & Facility\\
{[MJD]} & [hh:mm:ss.s] & [dd:mm:ss.s] & [$''$] & [hh:mm:ss.s] & [dd:mm:ss.s] & [$''$] & [hh:mm:ss.s] & [dd:mm:ss.s] & [$''$] &\\
\hline
\endfirsthead

\hline
& \multicolumn{3}{c}{Core} &  \multicolumn{3}{c}{Approaching ejection} & \multicolumn{3}{c}{Receding ejection} & \\
Date & RA & Dec & Error\tnote{a} & RA & Dec & Error\tnote{a} & RA & Dec & Error\tnote{a} & Facility\\
{[MJD]} & [hh:mm:ss.s] & [dd:mm:ss.s] & [$''$] & [hh:mm:ss.s] & [dd:mm:ss.s] & [$''$] & [hh:mm:ss.s] & [dd:mm:ss.s] & [$''$] &\\
\hline
\endhead

\hline
\multicolumn{6}{l}{{\textit{Continued on next page...}}} \\
\endfoot
\endlastfoot
58193.42 & 18:20:21.9384 & +07:11:7.182 & 0.004 & - & - & - & - & - & - & eMERLIN\\
58194.40 & 18:20:21.9386 & +07:11:7.172 & 0.003 & - & - & - & - & - & - & eMERLIN\\
58199.41 & 18:20:21.9391 & +07:11:7.182 & 0.008 & - & - & - & - & - & - & eMERLIN\\
58201.27 & 18:20:21.93867 & +07:11:7.169 & 0.001 & - & - & - & - & - & - & eMERLIN\\
58202.31 & 18:20:21.9385 & +07:11:7.166 & 0.006 & - & - & - & - & - & - & eMERLIN\\
58203.26 & 18:20:21.9384 & +07:11:7.168 & 0.004 & - & - & - & - & - & - & eMERLIN\\
58206.27 & 18:20:21.93858 & +07:11:7.168 & 0.001 & - & - & - & - & - & - & eMERLIN\\
58306.22\tnote{b} & - & - & - & 18:20:21.9382 & +07:11:07.157 & 0.003 & 18:20:21.93887 & +07:11:07.1785 & 0.0007 & VLBA\\
58308.98\tnote{c,d} & - & - & - & 18:20:21.9368 & +07:11:07.111 & 0.006 & - & - & - & eMERLIN\\
58308.98\tnote{c} & - & - & - & 18:20:21.9285 & +07:11:06.853 & 0.005 & - & - & - & eMERLIN\\
58310.02\tnote{c,d} & - & - & - & 18:20:21.9361 & +07:11:07.083 & 0.006 & - & - & - &  eMERLIN\\
58316.96\tnote{c} & - & - & - & 18:20:21.9145 & +07:11:06.308 & 0.005 & - & - & - & eMERLIN\\
58329.00\tnote{c} & - & - & - & 18:20:21.8780   & +07:11:05.230 & 0.006 & - & - & - & eMERLIN\\
58389.75 & 18:20:21.93 & +07:11:08.1 & 0.6 & 18:20:21.73  & +07:11:02.4  & 0.6 & - & - & - & MeerKAT\\
58396.70 & 18:20:21.91 & +07:11:07.6 & 0.6 & 18:20:21.71  & +07:11:01.3 & 0.6 & - & - & - & MeerKAT\\
58398.04 & 18:20:21.93 & +07:11:07.1 & 0.9 & 18:20:21.70 & +07:11:01.5 & 0.9 & - & - & - & VLA\\
58398.73\tnote{e} & 18:20:21.939 & 07:11:07.17 & 0.02 & 18:20:21.715 & +07:11:00.60 & 0.03 & - & - & - & eMERLIN\\
58399.99 & 18:20:21.94 & +07:11:07.3 & 0.9 & 18:20:21.75 & +07:11:00.9 & 0.9 & - & - & - & VLA\\
58402.85 & 18:20:22.00 & +07:11:07 & 1 & 18:20:21.73 & +07:11:00 & 1 & - & - & - & VLA\\
58403.66 & 18:20:21.92 & +07:11:07.9 & 0.6 & 18:20:21.68 & +07:11:01.2 & 0.6 & - & - & - & MeerKAT\\
58403.91 & 18:20:21.93 & +07:11:07.3 & 0.9 & 18:20:21.70 & +07:10:59.8 & 0.9 & - & - & - & VLA\\
58405.67 & 18:20:21.91 & +07:11:07.9 & 0.5 & 18:20:21.68 & +07:11:01.2 & 0.5 & - & - & - & MeerKAT\\
58405.90 & 18:20:21.93 & +07:11:07 & 1 & 18:20:21.66 & +07:11:01 & 1 & - & - & - & VLA\\
58410.62 & 18:20:21.94 & +07:11:08.0 & 0.6 & 18:20:21.67 & +07:11:01.3 & 0.6 & - & - & - & MeerKAT\\
58417.79 & 18:20:21.938 & +07:11:7.17 & 0.02 & - & - & - & - & - & - & eMERLIN\\
58419.73 & 18:20:21.939 & +07:11:7.17 & 0.02 & - & - & - & - & - & - & eMERLIN\\
58418.54 & 18:20:21.91 & +07:11:08.3 & 0.6 & 18:20:21.67 & +07:11:00.8 & 0.6 & - & - & - & MeerKAT\\
58418.85 & 18:20:21.96 & +07:11:06.8 & 0.8 & 18:20:21.72 & +07:10:59.6 & 0.8 & 18:20:22.10 & +07:11:10.7 & 0.8 & VLA\\
58425.50 & 18:20:21.91 & +07:11:08.1 & 0.6 & 18:20:21.65 & +07:11:01.1 & 0.6 & 18:20:22.02 & +07:11:12.3 & 0.7 & MeerKAT\\
58432.48 & 18:20:21.91 & +07:11:07.8 & 0.6 & 18:20:21.66 & +07:11:00.1 & 0.7 & 18:20:22.00 & +07:11:11.4 & 0.6 & MeerKAT\\
58435.67 & 18:20:21.90 & +07:11:07.4 & 0.5 & 18:20:21.67 & +07:10:59.8 & 0.7 & 18:20:22.01 & +07:11:11.1 & 0.5 & MeerKAT\\
58439.48 & 18:20:21.95 & +07:11:09.0 & 0.6 & 18:20:21.63 & +07:11:00 & 1 & 18:20:22.09 & +07:11:12.7 & 0.6 & MeerKAT\\
58440.90 & 18:20:21.94 & +07:11:07.2 & 0.2 & 18:20:21.69 & +07:10:59.2 & 0.2 & 18:20:22.06 & +07:11:11.0 & 0.3 & VLA\\
58446.45 & 18:20:21.89 & +07:11:06.5 & 0.8 & 18:20:21.55 & +07:10:57 & 1 & 18:20:22.03 & +07:11:12 & 1 & MeerKAT\\
58454.43 & 18:20:21.95 & +07:11:09 & 1 & 18:20:21.53 & +07:11:00 & 2 & 18:20:22.06 & +07:11:13.4 & 1 & MeerKAT\\
58473.68 & 18:20:21.94 & +07:11:07.0 & 0.3 & - & - & - & 18:20:22.08 & +07:11:11.6 & 0.3 & VLA\\
58479.64 & 18:20:21.93 & +07:11:07.1 & 0.4 & - & - & - & 18:20:22.10 & +07:11:12.3 & 0.4 & VLA\\
58484.75 & 18:20:21.96 & +07:11:07.0 & 0.3 & - & - & - & 18:20:22.09 & +07:11:12.0 & 0.3 & VLA\\
\hline
\end{longtable}
\begin{tablenotes}
\item [a] When only the core is detected the reported error is the statistical one reported by the CASA task IMFIT (RA and Dec error combined in quadrature). Otherwise it represents the uncertainty in the position along the angle connecting the components to the core (and the core to the components), further described in the Methods section.
\item [b] The position error reported for the VLBA observation is that along the jet axis, as described in the text. We use a core position measurement from the hard state, with a proper motion correction\supercite{gandhi2019}, when calculating the separation of the ejection components.
\item [c] This observation occurred when the source was not in the hard X-ray state, and as such the core was not detected.
\item [d] These observations were not included in our proper motion fits.
\item [e] While we detect the core in this observation, for the purpose of calculating the separation (\prettyref{sitab:allpos}) we use the bright core observation made on MJD 58201, see \prettyref{sitab:allflux}.
\end{tablenotes}
\end{ThreePartTable}

\clearpage

\begin{ThreePartTable}
\centering
\scriptsize
\begin{longtable}{ccccccccc}
\caption{\textbf{Flux evolution of the core, approaching ejection, and receding ejection from the 2018 outburst J1820.} Flux density of the approaching jet, receding jet, and core components for observations with eMERLIN, MeerKAT and the VLA. To calculate the flux density we use an unconstrained elliptical Gaussian and report the peak flux density. The error is the statistical one only, and was combined with a 5\% calibration error for calculations. Upper limits are $\textrm{3}\sigma$, although at early times when we cannot resolve the receding ejection component these may not reflect the true upper limit of the emitting region. We do not report upper limits before the launch date of the ejections. Dates report the observation mid-point.\label{sitab:allflux}}
\\
\hline
& \multicolumn{2}{c}{Core} &  \multicolumn{2}{c}{App. ejection} & \multicolumn{2}{c}{Rec. ejection} & & \\
Date & Flux density & Error & Flux density & Error & Flux density & Error & Frequency & Facility\\
{[MJD]} & [mJy] & [mJy] & [mJy] & [mJy]  & [mJy] & [mJy] & [GHz] &\\
\hline
\endfirsthead

\hline
& \multicolumn{2}{c}{Core} &  \multicolumn{2}{c}{App. ejection} & \multicolumn{2}{c}{Rec. ejection} & & \\
Date & Flux density & Error & Flux density & Error & Flux density & Error & Frequency & Facility\\
{[MJD]} & [mJy] & [mJy] & [mJy] & [mJy] & [mJy] & [mJy] & [GHz] &\\
\hline
\endhead

\hline
\multicolumn{6}{l}{{\textit{Continued on next page...}}} \\
\endfoot
\endlastfoot
58193.42 & 23.2 & 0.4 & - & - & - & - & 5.07 & eMERLIN\\
58194.40 & 26.6 & 0.4 & - & - & - & - & 5.07 & eMERLIN\\
58199.41 & 38 & 1 & - & - & - & - & 5.07 & eMERLIN\\
58201.27 & 56.7 & 0.8 & - & - & - & - & 5.07 & eMERLIN\\
58202.31 & 23 & 1 & - & - & - & - & 5.07 & eMERLIN\\
58203.26 & 26 & 1 & - & - & - & - & 5.07 & eMERLIN\\
58206.27 & 33.5 & 0.4 & - & - & - & - & 5.07 & eMERLIN\\
58308.98 & $<\textrm{0.08}$ & - & 0.24 & 0.02 & $<\textrm{0.08}$ & - & 5.07 & eMERLIN\\
58308.98 & $<\textrm{0.08}$ & - & 0.25 & 0.02 & $<\textrm{0.08}$ & - & 5.07 & eMERLIN\\
58310.02 & $<\textrm{0.13}$ & - & 0.52 & 0.04 & $<\textrm{0.13}$ & - & 5.07 & eMERLIN\\
58316.96 & $<\textrm{0.07}$ & - & 0.13 & 0.02 & $<\textrm{0.07}$ & - & 5.07 & eMERLIN\\
58329.00 & $<\textrm{0.10}$ & - & 0.35 & 0.04 & $<\textrm{0.10}$ & - & 5.07 & eMERLIN\\
58389.75 & 3.47 & 0.05 & 2.26 & 0.05 & $<\textrm{0.13}$ & - & 1.28 & MeerKAT\\
58396.70 & 11.8 & 0.1 & 2.0 & 0.1 & $<\textrm{0.19}$ & - & 1.28 & MeerKAT\\
58398.04 & 16.99 & 0.03 & 0.63 & 0.03 & $<\textrm{0.05}$ & - & 5.87 & VLA\\
58398.73 & 5.26 & 0.08 & 0.31 & 0.02 & $<\textrm{0.41}$ & - & 1.51 & eMERLIN\\
58399.99 & 7.46 & 0.05 & 0.50 & 0.04 & $<\textrm{0.06}$ & - & 6 & VLA\\
58402.85 & 5.12 & 0.03 & 0.33 & 0.03 & $<\textrm{0.08}$ & - & 6 & VLA\\
58403.66 & 2.62 & 0.04 & 1.06 & 0.04 & $<\textrm{0.11}$ & - & 1.28 & MeerKAT\\
58403.91 & 4.20 & 0.04 & 0.33 & 0.04 & $<\textrm{0.13}$ & - & 6 & VLA\\
58405.67 & 2.41 & 0.03 & 0.96 & 0.03 & $<\textrm{0.07}$ & - & 1.28 & MeerKAT\\
58405.90 & 3.59 & 0.05 & 0.28 & 0.05 & $<\textrm{0.12}$ & - & 6 & VLA\\
58410.62 & 1.52 & 0.06 & 0.77 & 0.06 & $<\textrm{0.016}$ & - & 1.28 & MeerKAT\\
58417.79 & 0.93 & 0.04 & $<\textrm{1.05}$ & - & $<\textrm{1.05}$ & - & 1.51 & eMERLIN\\
58419.73 & 1.15 & 0.03 & $<\textrm{0.21}$ & - & $<\textrm{0.21}$ & - & 1.51 & eMERLIN\\
58418.54 & 1.61 & 0.05 & 0.55 & 0.05 & $<\textrm{0.14}$ & - & 1.28 & MeerKAT\\
58418.85 & 2.49 & 0.03 & 0.17 & 0.01 & 0.15 & 0.03 & 6 & VLA\\
58425.50 & 1.15 & 0.04 & 0.41 & 0.04 & 0.36 & 0.04 & 1.28 & MeerKAT\\
58432.48 & 0.82 & 0.04 & 0.29 & 0.04 & 0.61 & 0.04 & 1.28 & MeerKAT\\
58435.67 & 0.75 & 0.02 & 0.25 & 0.02 & 0.55 & 0.02 & 1.28 & MeerKAT\\
58439.48 & 0.79 & 0.05 & 0.29 & 0.05 & 0.33 & 0.05 & 1.28 & MeerKAT\\
58440.90 & 1.162 & 0.007 & 0.071 & 0.007 & 0.22 & 0.007 & 6 & VLA\\
58446.45 & 0.36 & 0.05 & 0.25 & 0.05 & 0.35 & 0.05 & 1.28 & MeerKAT\\
58454.43 & 0.34 & 0.06 & 0.18 & 0.06 & 0.22 & 0.06 & 1.28 & MeerKAT\\
58473.68 & 0.138 & 0.008 & $<\textrm{0.02}$ & - & 0.13 & 0.008 & 6 & VLA\\
58479.64 & 0.153 & 0.008 & $<\textrm{0.03}$ & - & 0.10 & 0.008 & 6 & VLA\\
58484.75 & 0.147 & 0.008 & $<\textrm{0.02}$ & - & 0.10 & 0.008 & 6 & VLA\\
\hline
\end{longtable}
\end{ThreePartTable}

\clearpage

\begin{table}
\scriptsize
\caption{\textbf{Summary of our eMERLIN observations of MAXI J1820+070.}\label{sitab:emerlin}}
\vspace{5pt}
\noindent\makebox[\textwidth]{
\begin{threeparttable}
\begin{tabular}{ccccccc}
\hline
Date & Start time\tnote{a} & Start date\tnote{a} & Frequency & Obs. length\tnote{b} & Antennas\tnote{c} & RMS noise\tnote{d}\\
 & {[}UT{}] & {[}MJD{]} & {[}GHz{]} & {[}hrs.{]} &  & {[}$\mu\mathrm{Jy}\,\textrm{beam}^{-\mathrm{1}}${]}\\
\hline
16/03/2018 & 07:39:56.5 & 58193.31943 & 5.07 & 4.71 & Mk2, Kn, De, Pi & 319\\
17/03/2018 & 07:39:56.5 & 58194.31943 & 5.07 & 4.21 & Mk2, Kn, De, Pi, Da, Cm & 410\\
22/03/2018 & 07:09:56.5 & 58199.29859 & 5.07 & 4.83 & Mk2, Kn, De, Pi, Da, Cm & 766\\
24/03/2018 & 01:00:26.5 & 58201.04200 & 5.07 & 10.96 & Mk2, Kn, De, Pi, Da, Cm & 325\\
25/03/2018 & 02:53:02.5 & 58202.12019 & 5.07 & 9.08 & Mk2, Kn, De, Pi, Da, Cm & 1059\\
26/03/2018 & 01:07:56.5 & 58203.04720 & 5.07 & 10.27 & MK2, Ln, De, Pi, Da, Cm & 868\\
29/03/2018 & 01:07:56.5 & 58206.04720 & 5.07 & 10.83 & Mk2, Kn, De, Pi, Da & 217\\
09/07/2018 & 18:10:01.5 & 58308.75073 & 5.07 & 10.95 & Mk2, Kn, De, Pi, Da, Cm & 26\\
10/07/2018 & 20:03:01.5 & 58309.83546 & 5.07 & 8.95 & Mk2, Kn, De, Pi, Cm & 38\\
17/07/2018 & 17:01:00.5 & 58316.70906 & 5.07 & 11.95 & Kn, De, Pi, Da, Cm & 24\\
29/07/2018 & 20:05:01.5 & 58328.83685 & 5.07 & 7.95 & Mk2, Kn, De, Cm & 37\\
07/10/2018 & 12:01:02.0 & 58398.50073 & 1.51 & 10.95 & Mk2, Kn, De, Da, Cm & 69\\
26/10/2018 & 16:05:01.6 & 58417.67018 & 1.51 & 5.88 & Mk2, Kn, De, Pi, Da, Cm & 79\\
28/10/2018 & 13:31:02.0 & 58419.56323 & 1.51 & 7.95 & Mk2, Kn, De, Pi, Da, Cm & 42\\
\hline
\end{tabular}
\begin{tablenotes}
\item [a] Start time and Start data columns refer to the beginning on of the first scan on MAXI J1820.
\item [b] Observations length refers to the difference in time between the start of the first and end of the last scan on MAXI J1820. Roughly $\sim9$\% of this time was spent observing the interleaved phase calibrator.
\item [c] Mk2 = Mark II, Kn = Knockin, De = Defford, Pi = Pickmere, Da = Darnhall, Cm = Cambridge.
\item [d] RMS calculated from a region near the image phase centre. When the core was bright observations were dynamic range limited. 
\end{tablenotes}
\end{threeparttable}}
\end{table}

\clearpage

\begin{table}
\centering
\scriptsize
\caption{\textbf{Summary of our VLA observations of MAXI J1820+070.}\label{sitab:vla}}
\vspace{5pt}
\noindent\makebox[\textwidth]{
\begin{threeparttable}
\begin{tabular}{ccccccc}
\hline
Date & Start time & Start date & Frequency & Obs. length & Array config. & RMS noise\tnote{a}\\
& {[}UT{}] & {]}MJD{]} & {[}GHz{]} & {[}hrs.{]} &  & {[}$\mu\mathrm{Jy}\,\textrm{beam}^{-\mathrm{1}}${]}\\
\hline
07/10/2018 & 00:55:22 & 59398.03845 & 5.87 & 0.19 & D\tnote{b} & 17\\
08/10/2018 & 00:05:38 & 58399.00391 & 6.00 & 0.06 & D & 19\\
11/10/2018 & 20:47:47 & 58402.86652 & 6.00 & 0.06 & D & 26\\
12/10/2018 & 22:06:17 & 58403.92103 & 6.00 & 0.02 & D & 39\\
14/10/2018 & 21:58:18 & 58405.91549 & 6.00 & 0.02 & D & 40\\
27/10/2018 & 20:24:57 & 58418.85066 & 6.00 & 0.05 & D & 23\\
18/11/2018 & 21:31:22 & 58440.89678 & 6.00 & 0.60 & C\tnote{c} & 7\\
21/12/2018 & 16:22:22 & 58473.68220 & 6.00 & 0.31 & C & 8\\
27/12/2018 & 20:24:57 & 58479.85966 & 6.00 & 0.31 & C & 9\\
01/01/2019 & 18:14:32 & 58484.76009 & 6.00 & 0.31 & C & 8\\
\hline
\end{tabular}
\begin{tablenotes}
\item [a] RMS calculated from a region near the image phase centre.
\item [b] Maximum and minimum baseline length of 1.03 km and 0.035 km, respectively.
\item [c] Maximum and minimum baseline length of 3.4 km and 0.035 km, respectively.
\end{tablenotes}
\end{threeparttable}}
\end{table}

\clearpage

\begin{table}
\centering
\scriptsize
\caption{\textbf{Summary of our MeerKAT observations of MAXI J1820+070.}\label{sitab:meerkat}}
\vspace{5pt}
\noindent\makebox[\textwidth]{
\begin{threeparttable}
\begin{tabular}{cccccc}
\hline
Date & Start time\tnote{a} & Start date\tnote{a} & Frequency & Obs. length\tnote{b} & RMS noise\tnote{c}\\
& {[}UT{]} & {[}MJD{]} & {[}GHz{]} & {[}hrs.{]} & {[}$\mu\mathrm{Jy}\,\textrm{beam}^{-\mathrm{1}}${]}\\
\hline
28/09/2018 & 17:46:40.5 & 58389.74075 & 1.28 & 0.25 & 41\\
05/10/2018 & 16:33:42.5 & 58396.69008 & 1.28 & 0.24 & 72\\
12/10/2018 & 15:46:24.9 & 58403.65723 & 1.28 & 0.24 & 37\\
14/10/2018 & 15:15:56.8 & 58405.63607 & 1.28 & 1.71 & 24\\
19/10/2018 & 14:44:16.0 & 58410.61407 & 1.28 & 0.25 & 50\\
27/10/2018 & 12:49:17.8 & 58418.53423 & 1.28 & 0.25 & 45\\
03/11/2018 & 11:54:36.7 & 58425.49626 & 1.28 & 0.25 & 42\\
10/11/2018 & 11:26:18.6 & 58432.47660 & 1.28 & 0.25 & 44\\
13/11/2018 & 15:46:12.4 & 58435.65709 & 1.28 & 0.84 & 26\\
17/11/2018 & 11:26:41.8 & 58439.47687 & 1.28 & 0.25 & 53\\
24/11/2018 & 10:39:27.1 & 58446.44406 & 1.28 & 0.25 & 45\\
02/12/2018 & 10:05:03.5 & 58454.42018 & 1.28 & 0.25 & 57\\
\hline
\end{tabular}
\begin{tablenotes}
\item [a] Start time and Start data columns refer to the beginning on of the first scan on MAXI J1820.
\item [b] Observations length refers to the difference in time between the start of the first and end of the last scan on MAXI J1820. For observations of length 0.24 or 0.25 hours this was a single scan and thus the entire time was spent on source. For longer observations $\sim12$\% of this time was spent observing an interleaved phase calibrator.
\item [c] RMS calculated from a region near the image phase centre.
\end{tablenotes}
\end{threeparttable}}
\end{table}

\end{document}